\documentclass[a4paper,fleqn,10pt]{article}
\usepackage{amsmath}
\usepackage{amsfonts}
\usepackage{graphicx}
\usepackage[merge,numbers,compress]{natbib}
\usepackage{booktabs}
\usepackage{xcolor}
\usepackage{todonotes}
\usepackage{xspace}
\usepackage{dcolumn}
\usepackage{placeins}
\usepackage{makecell}
\usepackage[colorlinks=true,citecolor=blue!50!black,linkcolor=black]{hyperref}
\usepackage{caption}
\usepackage
[subrefformat=parens,position=top,skip=-15pt,margin=15pt,justification=justified,singlelinecheck=false]
{subcaption}
\usepackage{authblk}

\usepackage{mciteplus}
\mciteErrorOnUnknownfalse
\usepackage{geometry}
\usepackage{sectsty}
\usepackage{siunitx}
\usepackage{multirow}
\let\spreprint\empty
\newcommand{\preprint}[1]{\def\spreprint{\protect#1}}
\let\sinstitute\empty

\makeatletter
\renewcommand{\maketitle}{\begingroup
  \null\thispagestyle{empty}%
    \ifx\spreprint\empty
      \vskip 5ex
    \else
      \flushright\large\spreprint\vskip 2ex
    \fi
    \vskip 5ex
    \flushleft
      {\sffamily\bfseries\huge\@title}\vskip 2ex
      \@author\vskip 2ex
      \ifx\sinstitute\empty
      \else
        {\small\sinstitute}
      \fi
    \vskip 5ex
  \endgroup
}
\makeatother

\renewenvironment{abstract}{\begin{center}
  {\large\sffamily\bfseries Abstract: }
  \begin{minipage}[t]{0.75\textwidth}
}{\end{minipage}\end{center}\vskip 10ex}

\allsectionsfont{\normalfont\sffamily\bfseries}
\subsectionfont{\normalfont\sffamily\bfseries}
\subsubsectionfont{\normalfont\sffamily\bfseries}

\makeatletter
\DeclareRobustCommand*{\bfseries}{%
  \not@math@alphabet\bfseries\mathbf
  \fontseries\bfdefault\selectfont
  \boldmath
}
\makeatother

\def\beq{\begin{equation}}  
\def\eeq{\end{equation}}
\def\({\left(}
\def\){\right)}
\def\[{\left[}
\def\]{\right]}

\newcommand{\sherpa}{S\protect\scalebox{0.8}{HERPA}\xspace}
\newcommand{\apprentice}{A\protect\scalebox{0.8}{PPRENTICE}\xspace}

\newcommand{\comix}{C\protect\scalebox{0.8}{OMIX}\xspace}

\newcommand{\Caesar}{C\protect\scalebox{0.8}{AESAR}\xspace}
\newcommand{\Centauro}{C\protect\scalebox{0.8}{ENTAURO}\xspace}
\newcommand{\fastjet}{F\protect\scalebox{0.8}{AST}J\protect\scalebox{0.8}{ET}\xspace}
\newcommand{\rivet}{R\protect\scalebox{0.8}{IVET}\xspace}
\newcommand{\recola}{R\protect\scalebox{0.8}{ECOLA}\xspace}
\newcommand{\collier}{C\protect\scalebox{0.8}{OLLIER}\xspace}
\newcommand{\OpenLoops}{O\protect\scalebox{0.8}{PEN}L\protect\scalebox{0.8}{OOPS}\xspace}

\newcommand{\LHAPDF}{L\protect\scalebox{0.8}{HA}P\protect\scalebox{0.8}{DF}\xspace}
\newcommand{\HepMC}{H\protect\scalebox{0.8}{EP}MC\xspace}

\newcommand{\ahadic}{A\protect\scalebox{0.8}{HADIC++}\xspace}

\newcommand{\muR}{\ensuremath{\mu_{\text{R}}}}
\newcommand{\muF}{\ensuremath{\mu_{\text{F}}}}

\newcommand{\zcut}{\ensuremath{z_{\text{cut}}}}
\newcommand{\alphaS}{\alpha_\text{s}\xspace}

\newcommand{\NLO}{\text{NLO}\xspace}    
\newcommand{\NLL}{\text{NLL}\xspace}
\newcommand{\NLLp}{\ensuremath{\text{NLL}^\prime}\xspace}
\newcommand{\NLOpNLL}{\ensuremath{\NLO+\NLL}\xspace}
    
\newcommand{\NLOpNLLp}{\ensuremath{\NLOpNLL^\prime}\xspace}
\newcommand{\NLOpNLLpNP}{\ensuremath{\NLOpNLL^\prime+\text{NP}}\xspace}

\def\draftdate{\relax}
\def\mda{\relax}
\def\mua{\relax}
\def\mla{\relax}
\def\draft{
\def\thtystars{******************************}
\def\sixtystars{\thtystars\thtystars}
\typeout{}
\typeout{\sixtystars**}
\typeout{* Draft mode!
         For final version remove \protect\draft\space in source file *}
\typeout{\sixtystars**}
\typeout{}
\def\draftdate{\today}
\def\mua{\marginpar[\boldmath\hfil$\uparrow$]%
                   {\boldmath$\uparrow$\hfil}\color{black}%
                    \typeout{marginpar: $\uparrow$}\ignorespaces}
\def\mda{\color{red}\marginpar[\boldmath\hfil$\downarrow$]%
                   {\boldmath$\downarrow$\hfil}%
                    \typeout{marginpar: $\downarrow$}\ignorespaces}
\def\mla{\marginpar[\boldmath\hfil$\rightarrow$]%
                   {\boldmath$\leftarrow $\hfil}%
                    \typeout{marginpar: $\leftrightarrow$}\ignorespaces}
\def\Mua{\marginpar[\boldmath\hfil$\Uparrow$]%
                   {\boldmath$\Uparrow$\hfil}\color{black}%
                    \typeout{marginpar: $\uparrow$}\ignorespaces}
\def\Mda{\color{red}\marginpar[\boldmath\hfil$\Downarrow$]%
                   {\boldmath$\Downarrow$\hfil}%
                    \typeout{marginpar: $\downarrow$}\ignorespaces}
\def\Mla{\marginpar[\boldmath\hfil\textcolor{red}{$\Rightarrow$}]%
                   {\boldmath\textcolor{red}{$\Leftarrow $}\hfil}%
                    \typeout{marginpar: $\leftrightarrow$}\ignorespaces}
\overfullrule 5pt
\oddsidemargin 15mm
\marginparwidth 29mm
}

\usepackage{color}
\definecolor{darkblue}{rgb}{0,0,0.5}
\definecolor{darkred}{rgb}{0.5,0,0}
\definecolor{darkgreen}{rgb}{0,0.5,0}

\pdfoutput=1

\hypersetup{
  pdfauthor={Steffen Schumann et al.},
  pdftitle={(N)\NLOpNLLp accurate predictions for plain and groomed 1-jettiness in neutral current DIS at NNLO}
}

\preprint{MCNET-23-07\\IPPP/23/32}

\usepackage[symbol]{footmisc}

\author[1]{Max Knobbe\footnote[1]{Email: \texttt{max.knobbe@uni-goettingen.de}}}
\author[2]{Daniel~Reichelt\footnote[2]{Email: \texttt{daniel.reichelt@durham.ac.uk}}}
\author[1]{Steffen Schumann\footnote[5]{Email: \texttt{steffen.schumann@phys.uni-goettingen.de}}}

\affil[1]{Institut f{\"u}r Theoretische Physik, Georg-August-Universit{\"a}t G\"ottingen, Friedrich-Hund-Platz 1, 37077 G\"ottingen, Germany}
\affil[2]{Institute for Particle Physics Phenomenology, Department of Physics, Durham University,
  Durham DH1 3LE, United Kingdom}

\title{(N)\NLOpNLLp accurate predictions for plain and groomed 1-jettiness in neutral current DIS}
\begin{document}
\maketitle

\begin{abstract}
  The possibility to reanalyse data taken by the HERA experiments offers the chance to study
  modern QCD jet and event-shape observables in deep-inelastic scattering. To address this,
  we compute resummed and matched predictions for the 1-jettiness distribution
  in neutral current DIS with and without grooming the hadronic final state using the
  soft-drop technique.
  Our theoretical predictions also account for non-perturbative corrections from hadronisation
  through parton-to-hadron level transfer matrices extracted from dedicated Monte Carlo
  simulations with \sherpa. To estimate parameter uncertainties in particular for the
  beam-fragmentation modelling we derive a family of replica tunes to data from the HERA experiments.
  While NNLO QCD normalisation corrections to the NLO+NLL' prediction are numerically small,
  hadronisation corrections turn out to be quite sizeable. However, soft-drop grooming significantly
  reduces the impact of non-perturbative contributions.
  We supplement our study with hadron-level predictions from \sherpa based on the matching
  of NLO QCD matrix elements with the parton shower. Good agreement between the predictions
  from the two calculational methods is observed.
\end{abstract}

\clearpage
\vspace{10pt}
\noindent\rule{\textwidth}{1pt}
\tableofcontents
\noindent\rule{\textwidth}{1pt}
\vspace{10pt}

\section{Introduction}

Event shape observables offer great potential for detailed studies of the intriguing
dynamics of Quantum Chromodynamics (QCD), thereby providing insight into various strong
interaction phenomena. For example, they offer sensitivity to the strong coupling constant $\alpha_S$,
the colour charges of the QCD quanta, and parton density functions, when considering hadronic
initial state particles. Predictions for event shape distributions can be obtained
from fixed-order perturbation theory, all-orders resummation of logarithmically enhanced
contributions, as well as detailed particle-level simulations as provided by Monte Carlo
event generators. Accordingly, they form a rather unique testbed for a variety of theoretical
approaches, ranging from cutting-edge multi-loop calculations to detailed aspects in the
modelling of the non-perturbative parton-to-hadron transition.

Event shapes have played a central role in the QCD measurement program of past $e^+e^-$
collider experiments, see for instance~\cite{ALEPH:1990iba,OPAL:1990xiz,L3:1992btq,DELPHI:1999vbd,DELPHI:2003yqh}.
Also at hadron--hadron machines they are considered in studies of hadronic final states. Possibly even
more prominently, closely related jet-substructure observables have attracted enormous attention and
sparked the development of modern grooming and tagging techniques, see Ref.~\cite{Marzani:2019hun}
for a recent review. Also in deep-inelastic lepton--nucleon scattering experiments several
event shape variables have been measured~\cite{H1:1997hbl,H1:1999wfh,H1:2005zsk,ZEUS:1997nib,ZEUS:2002tyf,ZEUS:2006vwm}.
However, the LEP and HERA experiments phased
out in the years 2000 and 2007, respectively, such that later breakthroughs in calculational
methods and modern observable definitions have not yet been fully exploited.

Their complementarity and partially reduced complexity when compared to present day LHC
measurements, make the LEP and HERA data a real treasure for additional tests of our
theoretical understanding and simulation capabilities. In the past years a small number
of re-analyses of the LEP data have been published, see for
instance~\cite{Schieck:2012mp,ALEPH:2013htx,DELPHI:2014nkw,Fischer:2015pqa}.
Furthermore, there are efforts to provide open data sets that can directly be used by
the entire community~\cite{Badea:2019vey,Chen:2021uws}.

To open the treasure chest of their large data set for modern QCD studies the HERA H1
collaboration has recently started to publish a series of new, fascinating measurements
that allow one to confront contemporary state-of-the-art predictions with precise DIS
data. Besides their relevance for benchmarking our present day tools, such analyses build
an important stepping stone towards future electron--hadron colliders like the EIC at
BNL~\cite{Accardi:2012qut,AbdulKhalek:2021gbh} or the LHeC at
CERN~\cite{LHeCStudyGroup:2012zhm,LHeC:2020van}.

We here compile predictions for the 1-jettiness event shape in the Breit frame~\cite{Kang:2013nha},
that is equivalent to the well known thrust variable \cite{Antonelli:1999kx}, for the HERA kinematics,
\emph{i.e.}\ lepton--proton collisions at $\sqrt{s}=319\;\text{GeV}$. Furthermore, we consider
grooming of the hadronic final states based on the soft-drop method prior to the observable evaluation.
We derive differential distributions for groomed and ungroomed $\tau^b_1$ differential in the photon
virtuality $Q^2\in[150,20000]\;\text{GeV}^2$, and the events inelasticity $y\in[0.05,0.94]$.
  We perform Monte Carlo simulations with the \sherpa generator based on next-to-leading-order (NLO)
  matrix elements for the one- and two-jet final states matched to the parton shower
  and hadronised using \sherpa's new cluster fragmentation model~\cite{Chahal:2022rid}.
  To estimate the hadronisation modelling uncertainties in particular related to the
  beam remnant fragmentation we derive a set of replica tunes~\cite{Knobbe:2023njd} to a selection
  of DIS measurements from the H1 and ZEUS experiments.

  Furthermore, we compute resummed predictions at next-to-leading-logarithmic (NLL) accuracy in
  the observable value based on the implementation of the \Caesar resummation
  formalism~\cite{Banfi:2004yd} in the \sherpa framework~\cite{Gerwick:2014gya}. These get
  matched to the NNLO QCD result for the inclusive DIS process and the NLO matrix elements for
  the two-jet channel. For the NNLO QCD corrections we rely on an implementation in \sherpa
  presented in~\cite{Hoche:2018gti}. {This results in predictions of \NLOpNLLp accuracy
    for the actual event-shape distributions,  while we achieve NNLO precision for the total event rate.
    In consequence, we refer to our predictions as being (N)\NLOpNLLp accurate.} To account for
  non-perturbative corrections we derive parton-to-hadron level transfer matrices differential in the
  event shape variables that we extract from particle level simulations with \sherpa~\cite{Reichelt:2021svh},
  thereby also accounting for the cluster-model parameter uncertainties through the set of
  replica tunes to HERA data.

  Our calculations are targeted on an upcoming measurement by the H1 experiment, for that preliminary
  results have recently been presented~\cite{Hessler:2021usr,Hessler:2021ubp}. Results based on simulations
  with \sherpa in a similar fiducial phase space have been compared to data from jet-substructure
  observables in neutral current DIS in~\cite{H1:2023fzk}. Our study extends earlier work on the
  simulation of DIS events with \sherpa~\cite{Carli:2010cg}. Furthermore, this is the first
  time we include NNLO QCD correction in resummation calculations with \sherpa.

  The article is organised as follows: in Sec.~\ref{sec:definitions} we introduce
  the considered observables and define the fiducial phase space used in
  our study of the hadronic final states produced in $ep$ collisions at HERA.
  In Sec.~\ref{sec:sherpasetup} we describe the setup used to simulate DIS events
  with \sherpa as well as the tuning of its beam-fragmentation parameters. In Sec.~\ref{sec:res_setup} we present our framework to compile
  (N)\NLOpNLLp predictions, based on the implementation of the \Caesar formalism in \sherpa.
  Here, we also present our approach to treat non-perturbative corrections based on
  transfer matrices extracted from MC simulations, see Sec.~\ref{sec:res_review}. We present our
  final (N)\NLOpNLLpNP results in Sec.~\ref{sec:results}, alongside
  with MC predictions from \sherpa. We compile our conclusions and give an outlook
  in Sec.~\ref{sec:conclusions}.

\section{Phase space and observable definition}\label{sec:definitions}

We consider deep-inelastic scattering (DIS) of leptons with momentum $p$ of
off protons with momentum $P$ at HERA energies, \emph{i.e.}\ $E_l=27.6\;\text{GeV}$ and
$E_p=920\;\text{GeV}$, resulting in a centre-of-mass energy of $\sqrt{s}=319\;\text{GeV}$.
Denoting the outgoing lepton momentum as $p^\prime$, we define the momentum difference,
at LO carried by the virtual photon, as
\begin{equation}
  q = p - p^\prime \equiv (0,0,0,-Q)~,
\end{equation}
where the last equivalence defines the Breit frame, which we will assume
whenever frame-specific formulae are given. We also introduce the usual Bjorken
variable $x_B$ and inelasticity $y$
\begin{align}
  x_B &= \frac{Q^2}{2 P\cdot q}\,,\\
  y   &= \frac{P\cdot q}{P\cdot p}~.
\end{align}
We consider events with $150 < Q^2/\text{GeV}^2 < 2\cdot 10^4$ and $0.05 < y < 0.94$.
No other cuts are applied, but we have studied 1-jettiness in smaller bins of $Q^2$
and $y$, and will only discuss a selection
of results here\footnote{Results over the full range of $Q^2, y$ in several bins of
both variables are available upon request.}.

We take into account all final state particles apart from the outgoing lepton
for the calculation of event-shape variables. We study a well known observable, referred
to as thrust $\tau_Q$~\cite{Antonelli:1999kx} or alternatively 1-jettiness $\tau^b_1$~\cite{Kang:2013nha}.
Several equivalent definitions exist in the literature. For concreteness we
define it by dividing the event into a current hemisphere $\mathcal{H}_C$ and a
beam hemisphere $\mathcal{H}_B$. Working in the Breit frame, we can introduce
two reference vectors
\begin{equation}
  n_\pm = (1,0,0,\pm1)
\end{equation}
and denote the hemispheres according to the final state particles momentum fractions
along those,
\begin{align}
  \mathcal{H}_C = \{p_i : p_i \cdot n_+ > p_i \cdot n_-\}\quad\text{and}\quad \mathcal{H}_B = \{p_i : p_i \cdot n_+ < p_i \cdot n_-\}~.
\end{align}
We can now define thrust as the sum of the longitudinal momentum components of all particles
in the current hemisphere. As we prefer to work with an observable that vanishes
in the soft limit, {\emph{i.e.} the limit where all final state partons apart
from the struck quark have arbitrarily small momenta}, we ultimately use
\begin{equation}\label{eq:tau_def}
  \tau = 1 - \frac{2}{Q} \sum_{p_i \in \mathcal{H}_C} p_i^z\,.
\end{equation}
Despite this definition only summing over one of the hemispheres, thrust, \emph{i.e.}\
1-jettiness, is actually sensitive to emissions anywhere in the event, and indeed is a
global event shape in the sense of \emph{e.g.} \cite{Banfi:2004yd}. Note this statement
depends on the precise definition, including the normalisation factor here given by $Q/2$,
that differs in the thrust variant we use for tuning in the following.

In addition we study 1-jettiness calculated based on events that have been groomed of
soft wide-angle radiation. Soft-drop grooming was first introduced in
\cite{Larkoski:2014wba} as a jet substructure technique, including as a special
case the modified Mass Drop Tagger \cite{Butterworth:2008iy, Dasgupta:2013ihk}.
It has since been generalised and applied also to jets at lepton colliders
\cite{Baron:2018nfz, Chen:2021uws} and event shapes at both lepton
\cite{Baron:2018nfz, Marzani:2019evv} and hadron \cite{Baron:2020xoi}
colliders. A version applicable to DIS was proposed in \cite{Makris:2021drz},
based on the \Centauro jet algorithm \cite{Arratia:2020ssx}, that accounts for the
forward-backward asymmetry when considering the Breit frame. This sequential
cluster algorithm is based on the distance measure between particles with momenta
$p_i, p_j$
\begin{align}
  d_{ij} &= (\Delta\bar{z}_{ij})^2 + 2\bar{z}_i\bar{z}_j(1-\cos\Delta\phi_{ij})\,,\label{eq:centauro_def}\\
  \text{with}\;\;\bar{z}_i &= 2\sqrt{1+\frac{q\cdot p_i}{x_B P\cdot p_i}}\quad\text{and}\quad ~\Delta\bar{z}_{ij}=\bar{z}_{i}-\bar{z}_{j}\,.\label{eq:centauro_def_zb}
\end{align}
Note that \cite{Arratia:2020ssx} discusses more general functional
forms of the distance measure, while we concentrate here on the definition given
in \cite{Makris:2021drz}. As in all other soft-drop grooming methods the
objects of interest, in this case the full event, are first clustered according
to this sequential algorithm, and then the reverse clustering history is considered.
The last cluster step is undone, and the softness of the softer of the two branches
is evaluated. For the DIS case, \cite{Makris:2021drz} suggests to use
\begin{equation}\label{eq:dis_sd_z}
  z_i = \frac{P\cdot p_i}{P\cdot q}
\end{equation}
as a measure for softness. The formal soft-drop criterion then reads
\begin{equation}\label{eq:soft_drop_inequ}
  \frac{\min[z_i,z_j]}{z_i+z_j} > \zcut~,
\end{equation}
with $\zcut$ the grooming parameter.
If this is satisfied, \emph{i.e.} both branches are classified as hard, the
algorithm terminates. Otherwise the softer branch (with smaller $z$) is
dropped, and the procedure is repeated with the harder branch. This iteration
stops when either Eq.~\eqref{eq:soft_drop_inequ} is satisfied, or there is only
one particle left in the hard branch such that no further unclustering is possible.

We finally recalculate 1-jettiness, using Eq.~\eqref{eq:tau_def} but restricting the
sum to particles in the current hemisphere that have not been dropped during grooming,
thereby considering variable values for $\zcut$.

\section{DIS Monte Carlo simulations with \sherpa}\label{sec:sherpasetup}

We derive hadron-level predictions for the DIS event shapes using a pre-release
version of \sherpa-3.0~\cite{Sherpa3.0.beta}, that will supersede the current \sherpa-2.2
series~\cite{Sherpa:2019gpd}. This major release features extended physics-modelling
capabilities, including, for example, the automated evaluation of electroweak (EW) corrections
at the one-loop order~\cite{Schonherr:2017qcj,Schonherr:2018jva,Brauer:2020kfv} or
in the Sudakov approximation~\cite{Bothmann:2020sxm,Bothmann:2021led}, a complete
reimplementation of the cluster hadronisation model~\cite{Chahal:2022rid}, as well
as an improved user interface based on {\sc{Yaml}}~\cite{yaml}. To analyse our
simulated event samples we employ the \rivet analysis package~\cite{Bierlich:2019rhm}.
For jet clustering we use the \Centauro plugin~\cite{Arratia:2020ssx} within the \fastjet
framework~\cite{Cacciari:2011ma}.

\subsection{MEPS@NLO predictions for DIS}
\label{sec:dissimulation}
The basics of simulating DIS processes by merging parton-shower evolved
higher-multiplicity tree-level matrix elements within the \sherpa framework have been
presented in~\cite{Carli:2010cg}. We here lift this to next-to-leading order (NLO)
accurate QCD matrix elements. To this end, we consider the massless single and dijet production
channels in neutral current DIS at NLO, and three- and four-jets at leading order (LO),
\emph{i.e.}\
\begin{equation}\label{eq:MEPS_MEs}
  e^-p\to e^-+1,2\,j\,@\,\text{NLO}+3,4\,j\,@\,\text{LO},
\end{equation}
where we consider $u,d,s$ quarks to be massless and add additional LO processes for the remaining massive quarks.
The massless and massive channels get matched to the \sherpa Catani--Seymour dipole shower~\cite{Schumann:2007mg} and
merged according to the MEPS@NLO~\cite{Hoeche:2012yf} and MEPS@LO~\cite{Hoeche:2009rj} truncated shower formalism,
respectively. The contributing one-loop amplitudes are obtained from \OpenLoops~\cite{Buccioni:2019sur},
that employs the \collier library~\cite{Denner:2016kdg} for the evaluation of tensor and
scalar integrals. All tree-level matrix elements are provided by \comix~\cite{Gleisberg:2008fv}, and PDFs are
obtained from LHAPDF \cite{Buckley:2014ana}.

To determine the perturbative scales entering the calculation, the final states of the multi-parton
final states get clustered to a two-to-two core process~\cite{Hoeche:2009rj}. For the reconstructed
core the factorisation, renormalisation, and parton shower starting scale are set to
\begin{eqnarray}\label{eq:scalesDIS}
  \mu_{\text{F}}=\mu_{\text{R}}=\mu_\text{Q}:=\mu_{\text{core}}\,.
\end{eqnarray}
For jet-associated DIS three configurations need to be distinguished~\cite{Carli:2010cg}:

\begin{enumerate}
  \item[(i)] virtual photon exchange, \emph{i.e.}\ $ej\to ej$, where $\mu_{\text{core}}^2=Q^2$,
  \item[(ii)] interaction of the virtual photon with a QCD parton,
    \emph{i.e.}\ $\gamma^*j\to j_1j_2$, with $\mu^2_{\text{core}}=m_{\perp,1}
    m_{\perp,2}$ defined as the product of the two jet transverse masses
    $m_{\perp,i} = \sqrt{m^2_i + p_{\perp,i}^2}$ relative to the beam axis,
  \item[(iii)] and pure QCD channels, \emph{i.e.}\ $jj\to jj$, where
    $\mu^2_{\text{core}}=-\frac{1}{\sqrt{2}}\left(s^{-1}+t^{-1}+u^{-1}\right)^{-1}$
    is a scaled harmonic mean of the Mandelstam variables $s,t,u$.
\end{enumerate}

Beyond the core process, the arguments of the strong-coupling factors are determined
by the clustering algorithm~\cite{Hoeche:2009rj}. The merging-scale parameter, separating the different jet-multiplicity contributions, is dynamically set to
\begin{equation}
  Q_{\text{cut}} = \frac{\bar{Q}_{\text{cut}}}{\sqrt{1+\bar{Q}^2_{\text{cut}}/Q^2}}\,,\quad\text{using}\quad \bar{Q}_{\text{cut}}=5\,\text{GeV}\,.
\end{equation}
As parton density functions we use the NNLO PDF4LHC21\_40\_pdfas set~\cite{PDF4LHCWorkingGroup:2022cjn} with
$\alpha_S(M^2_Z)$=0.118.

To estimate perturbative uncertainties, we consider 7-point variations of the factorisation
($\mu_F$) and renormalisation ($\mu_R$) scales in the matrix element and the parton shower
that get evaluated on-the-fly~\cite{Bothmann:2016nao}, \emph{i.e.}\
\begin{equation}\label{eq:7pointvar}
  \{(\tfrac{1}{2}\muR, \tfrac{1}{2}\muF)$, $(\tfrac{1}{2}\muR,\muF),(\muR,\tfrac{1}{2}\muF)$, $(\muR,\muF)$, $(\muR,2\muF),(2\muR,\muF)$, $(2\muR,2\muF)\}\,.
\end{equation}
The resummation scale $\mu_Q$ we keep fixed. {The final uncertainty estimate is
derived by forming an envelope of all variations.}

\subsection{Tuning the beam fragmentation model against HERA data}\label{sec:tuning}

Ref.~\cite{Chahal:2022rid} presented a new cluster fragmentation model for \sherpa that will
be used in \sherpa-3, superseding the old cluster model described in~\cite{Winter:2003tt},
that was used in the \sherpa-1.X~\cite{Gleisberg:2008ta} and \sherpa-2.X~\cite{Sherpa:2019gpd}
released. A particular feature of the new implementation is a specific treatment of the
fragmentation of hadronic clusters that contain beam remnant particles. To calibrate the
corresponding model parameters we performed dedicated tunes using HERA data for hadronic final
state observables in neutral current DIS.

Broadly speaking, a cluster hadronisation simulation features two basic components, a
cluster-formation and a cluster-decay model~\cite{Marchesini:1983bm,Webber:1983if}. Based on
the pre-confinement property of QCD~\cite{Amati:1979fg}, finite mass colour neutral mesonic
and baryonic clusters can be formed from the final state of a parton shower evolution of a hard scattering event.
These primary clusters are then subject to an iterative fission process that ultimately
results in the transition to known hadronic resonances, whose decays can be treated by
a dedicated package. Both elements of the hadronisation model introduce sets of parameters
that need to be carefully adjusted by comparing model predictions and measurements for
suitable observables, a process commonly known as tuning.

In Ref.~\cite{Knobbe:2023njd} the free model parameters were calibrated against hadronic
observables measured in electron--positron annihilation experiments. However, in leptonic
collisions the beam fragmentation modelling is not probed and the corresponding
parameters remained unconstrained. This affects in particular the parametrisation of
the decay of clusters that contain a remnant particle of an incident hadron, \emph{e.g.}\ a
(anti-)quark and (anti-)diquark from the break-up of the incoming proton in DIS. We consider
the two-body decay of a beam cluster with flavours $f_1$ and $\bar{f}_2$,
where a (di)quark-flavour pair $f\bar{f}$ is drawn from the vacuum, resulting in
\begin{equation}
{\cal{C}}[f_1\bar{f}_2]\to {\cal{C}}_1[f_1\bar{f}]\;{\cal{C}}_2[f\bar{f}_2]\,.
\end{equation}
To fix the kinematics of the two-body decay in the rest frame of ${\cal{C}}$, the
absolute value of the transverse momentum of the decay products ${\cal{C}}_1$ and
${\cal{C}}_2$ is selected according to a Gaussian distribution ${\cal{N}}(0,k^2_{T,0}/2)$ that is
truncated at the parton-shower cut-off $p_{T,\text{min}}$, \emph{i.e.}\
\begin{equation}
{\cal{P}}(k_T) \propto \exp\left(-k_T^2/k_{T,0}^2\right) \Theta(p^2_{T,\text{min}}-k_T^2)\,.
\end{equation}
The parameter $k_{T,0}$ is thereby considered as independent of the incident cluster type.
The direction of the two-component $\vec{k}_T$ is picked uniformly in the transverse plane,
with $f_1$ and $\bar{f}_2$ pointing along the positive and negative $z$-axis, respectively.
This leaves one to fix the longitudinal momentum fractions $z^{(1),(2)}$ with respect to the
light-like vectors $n^\mu_\pm=(1,0,0,\pm 1)$. For the case of a beam-remnant cluster, still
working in its rest frame, these are distributed according to
\begin{equation}\label{eq:frag_zdist}
  {\cal{P}}(z) \propto z^{\alpha_B}(1-z)^{\beta_B}\cdot\exp\left\{-\gamma_B\frac{1}{z}\left(\frac{k^2_T+(m_{f_1}+m_{\bar{f}_2})^2}{k^2_{T,0}}\right)\right\}\,.
\end{equation}
Note the similarity to the symmetric Lund string fragmentation function~\cite{Andersson:1983ia}.

This results in the four-momenta of the decay products being given by
\begin{eqnarray}
  p^\mu_{{\cal{C}}_1} &=& \frac{m_{{\cal{C}}}}{2}\left(z^{(1)}n^\mu_++(1-z^{(2)})n^\mu_-\right)+k^\mu_T\,,\\
  p^\mu_{{\cal{C}}_2} &=& \frac{m_{{\cal{C}}}}{2}\left((1-z^{(1)})n^\mu_++z^{(2)}n^\mu_-\right)-k^\mu_T\,.
\end{eqnarray}
According to Eq.~\eqref{eq:frag_zdist} the relevant free parameters
specifically steering the decays of beam clusters are $\alpha_B$,
$\beta_B$, and $\gamma_B$. To calibrate those we performed dedicated
tunes based on a variety of hadronic observables measured by the HERA
experiments H1 and ZEUS. The remaining hadronisation parameters are
set according to the LEP data tune described in Ref.~\cite{Knobbe:2023njd}.

We employ the \apprentice tuning tool \cite{Krishnamoorthy:2021nwv}, with
reference data for DIS analyses at centre of mass energies of
$\sqrt{s} = \SI{300}{\giga\eV}$, \emph{i.e}. lepton energies of
$\SI{27.5}{\giga\eV}$ and proton energies of $\SI{820}{\giga\eV}$.
The tuning requires an initial set of Monte Carlo runs, that are then used
to generate a polynomial, bin-wise approximation of the Monte Carlo response
with respect to changes in the hadronisation-model parameters. The predictions
for the grid points are generated using the calculational setup described in
Sec.~\ref{sec:dissimulation}.

The selection of observables considered for the tuning includes classic
variables sensitive to hadronisation. In particular, we use event-shape distributions
like thrust and jet broadening~\cite{H1:2005zsk}, energy flows and charged particle
spectra~\cite{H1:1999dbb,H1:1994vjx} and multiplicities~\cite{ZEUS:1995red,H1:1996ovs},
as well as quark fragmentation functions~\cite{H1:1995cqf,H1:1997mpq}. Further
details on the used analyses and observables are provided in App.~\ref{app:tuning-details}.

Given we consider model parameters newly introduced that have not been
tuned before, we have little prior knowledge about their preferred values and
thus need to start out with rather wide parameter ranges. To narrow these down, we
make an initial pass to get a rough idea of the relevant regions. The corresponding
ranges are outlined in Tab.~\ref{tab:tuning_ranges}. For a second run we restrict the
tuning ranges using the results of the exploration run, resulting in an iterative
procedure to further narrow down the considered parameter intervals. The initial run,
with largely unconstrained parameter values also serves the purpose of filtering out
the most sensitive observables from the considered analyses. Observables or observable
regions that remain unchanged under the variation of the tuning parameters are
not suited for the following tunes and therefore dropped.

\begin{figure}
  \includegraphics[width=.32\textwidth]{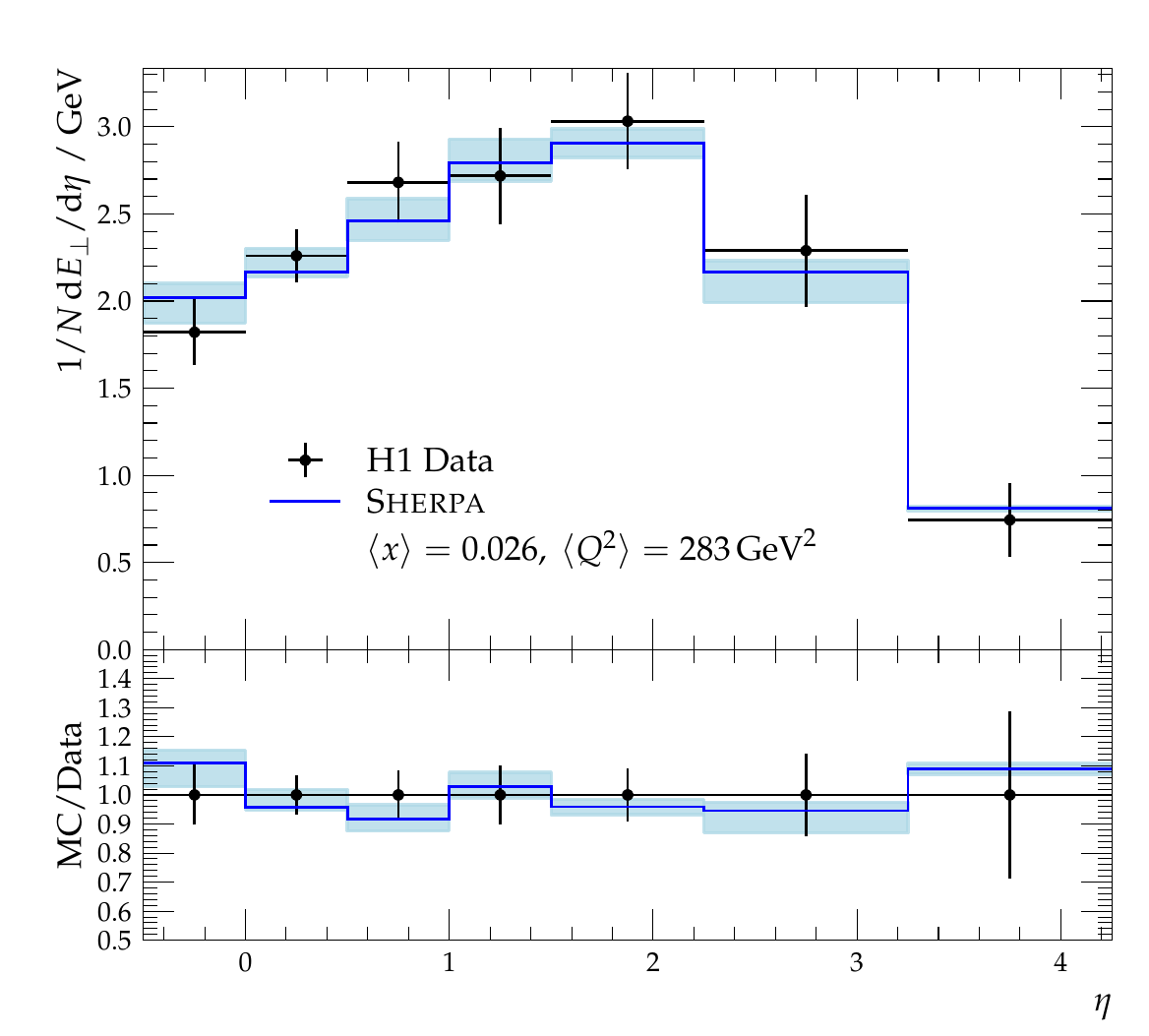}
  \includegraphics[width=.32\textwidth]{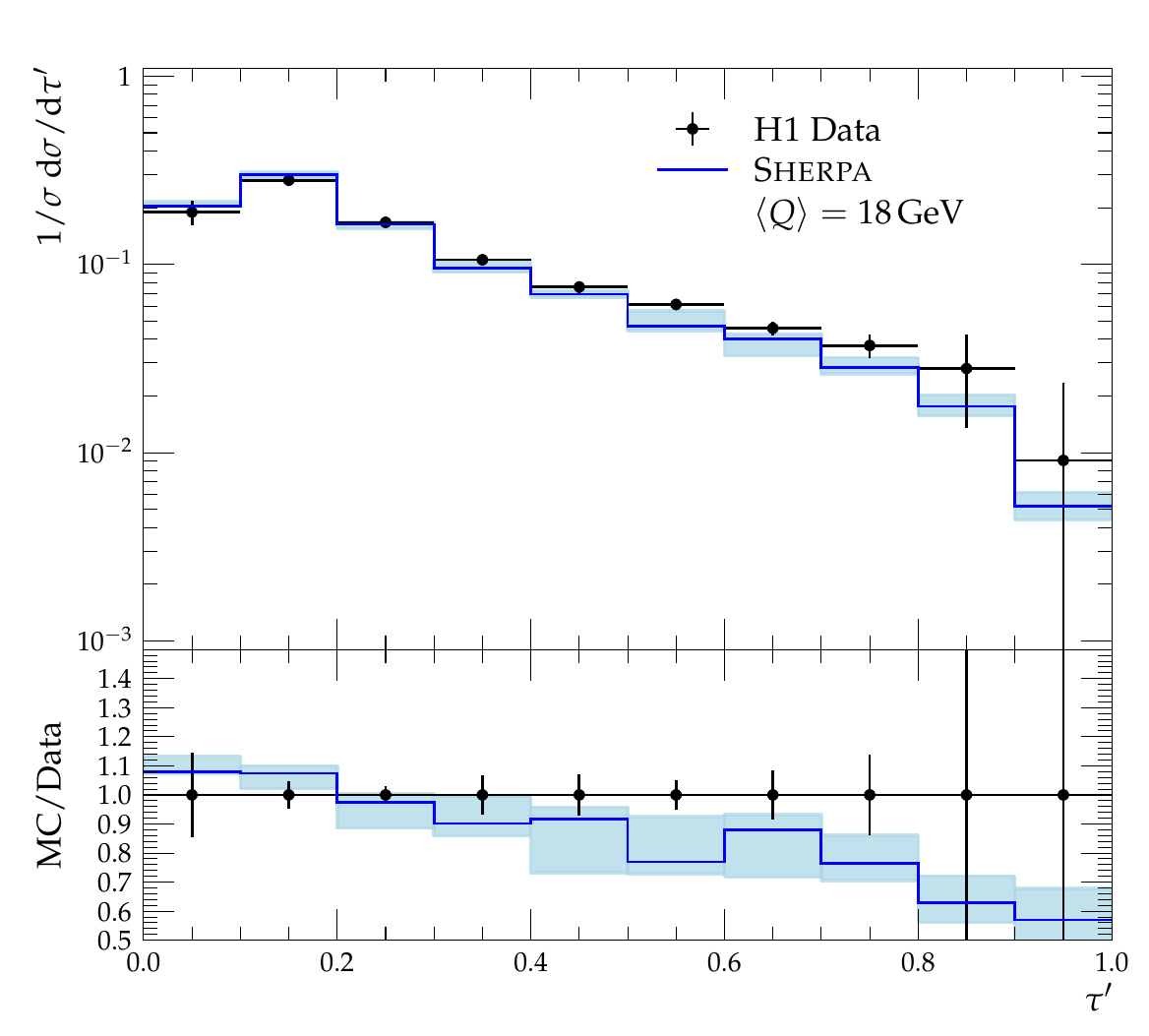}
  \includegraphics[width=.32\textwidth]{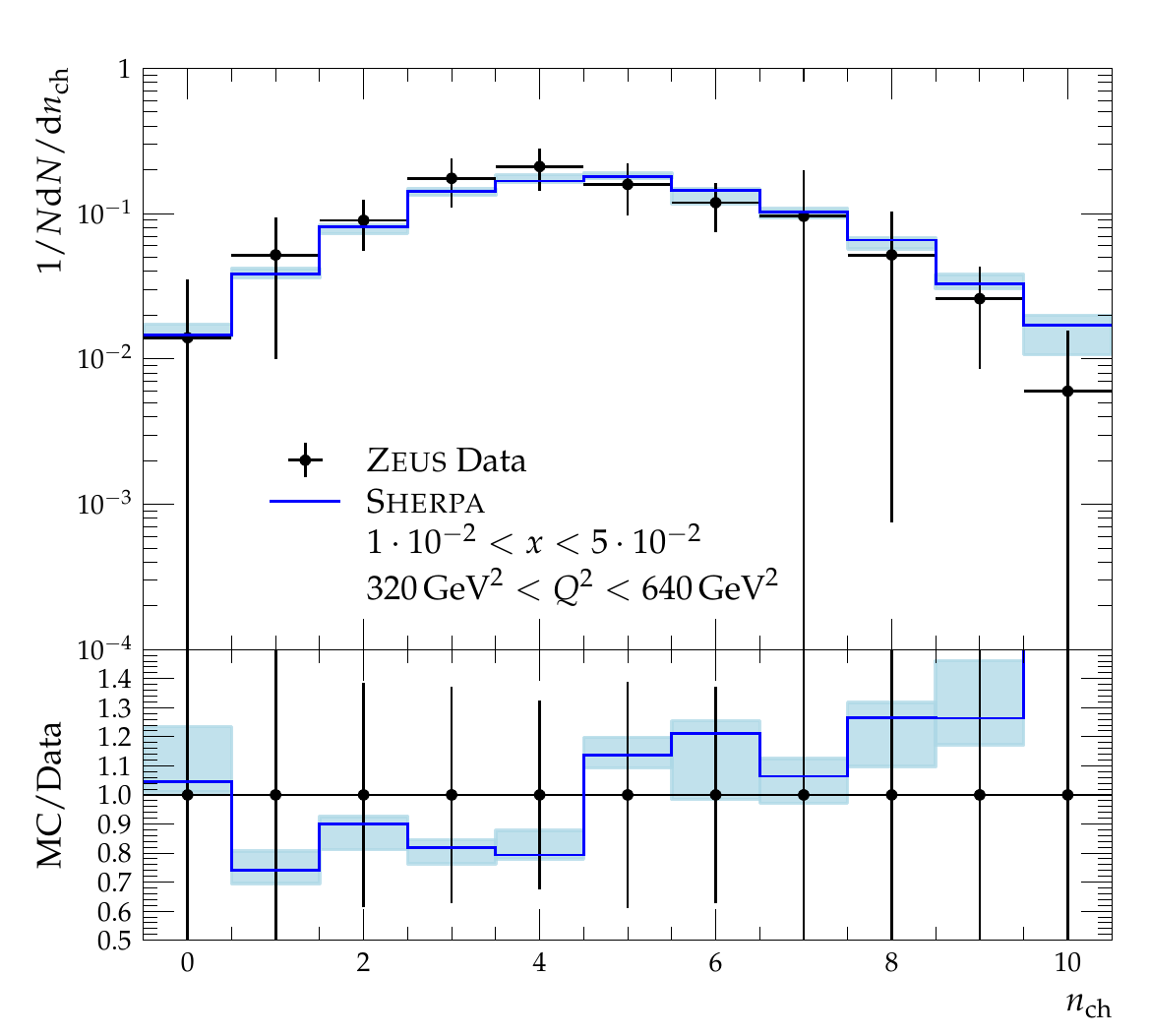}
  \caption{\sherpa predictions for the hadronisation tune, for observables
    measured by the H1 and ZEUS experiments at $\sqrt{s} = \SI{296}{\giga\eV}$. Shown
    is the transverse energy flow (left) \cite{H1:1999dbb}, thrust $\tau'$ (center) \cite{H1:2005zsk}
    and the charged particle multiplicity $n_\mathrm{ch}$ (right) \cite{ZEUS:1995red}. Note,
    the statistical uncertainties of the simulated data is small compared to the
    non-perturbative tuning uncertainties indicated by the blue band.}
  \label{fig:tuning_plots}
\end{figure}

Similar to the procedure described in Ref.~\cite{Knobbe:2023njd}, we
generate a set of equivalent tunes that only differ by the Monte Carlo
runs used to construct the polynomial approximations as
described above. The tunes are thus fully equivalent and can be used
to estimate the non-perturbative model-parameter uncertainties as
illustrated in Fig.~\ref{fig:tuning_plots} for a selection of data from the
HERA experiments. We call these alternative parameter sets replica tunes. To reflect
the uncertainty associated with the three beam-fragmentation parameters we here
consider seven such replicas, \emph{cf.} Tab.~\ref{tab:tuning_ranges} for the
resulting uncertainty variations.

\begin{table}
  \centering
  \begin{tabular}{c c c c c}
    \toprule
    parameter & parameter tag & tuning range & central tune & uncertainty variation\\
    \midrule
    $\alpha_B$ & {\tt ALPHA\_B} & [-1, 20]  & 14.2 & [13.9, 14.8]\\
    $\beta_B$  & {\tt BETA\_B } & [0.5, 4]  & 1.59 & [1.14, 1.60]\\
    $\gamma_B$ & {\tt GAMMA\_B} & [1, 20]   & 8.11 & [8.06, 9.47]\\
    \bottomrule
  \end{tabular}
  \caption{\ahadic model parameters considered in the tuning. Quoted are
    the initial parameter interval, the obtained central-tune value, and
    uncertainty ranges extracted from 7 replica tunes.}
  \label{tab:tuning_ranges}
\end{table}

\section{(N)\NLOpNLLp resummation for 1-jettiness in DIS}\label{sec:res_setup}

The 1-jettiness observable considered here is equivalent to thrust in DIS, which
has originally been resummed at NLL accuracy in \cite{Antonelli:1999kx,
  Dasgupta:2002dc}. The more general $n$-jettiness~\cite{Stewart:2010tn,Jouttenus:2011wh}
was suggested for lepton--hadron collisions in \cite{Kang:2012zr}, and has been resummed
to NNLL accuracy \cite{Kang:2013wca}. For 1-jettiness, analytic fixed order results at LO have
been presented in \cite{Kang:2014qba}, and the NLL calculation has been matched
to fixed order at NLO accuracy in \cite{Kang:2013lga}. The resummed calculations
in this formalism for event shapes in DIS were extended to N$^3$LL in
\cite{Kang:2015swk}.
Grooming for DIS has first been suggested in \cite{Makris:2021drz} based on jets
defined with the \Centauro jet algorithm \cite{Arratia:2020ssx}. The same
Ref.~\cite{Makris:2021drz} also provided NNLL results for both 1-jettiness and
jet mass after soft drop grooming. Non-perturbative corrections have there been
modelled through a two-parameter shape function~\cite{Frye:2016aiz,Hoang:2019ceu}.
To our knowledge there are no published results studying these observables including
matching to fixed order or using a fixed order calculation alone.

\subsection{NLL resummation in the \Caesar approach}\label{sec:res_review}

To perform the NLL resummation of logarithms $L$ of event shapes in
DIS we use the implementation of the \Caesar formalism~\cite{Banfi:2004yd}
available in the \sherpa framework~\cite{Gerwick:2014gya,Baberuxki:2019ifp}.
For a recursive infrared and collinear (rIRC) safe observable, the cumulative cross section
for observable values up to $v = \exp(-L)$ can be expressed to all orders, in
general as a sum over partonic channels $\delta$, as follows:
\begin{equation}\label{eq:CAESAR}
  \begin{split}
    \Sigma_\mathrm{res}(v) &= \sum_\delta \Sigma_\mathrm{res}^\delta(v)\,,\,\,\text{with} \\
    \Sigma_\mathrm{res}^\delta(v) &= \int d\mathcal{B_\delta}
    \frac{\mathop{d\sigma_\delta}}{\mathop{d\mathcal{B_\delta}}} \exp\left[-\sum_{l\in\delta}
      R_l^\mathcal{B_\delta}(L)\right]\mathcal{P}^{\mathcal{B}_\delta}(L)\mathcal{S}^\mathcal{B_\delta}(L)\mathcal{F}^\mathcal{B_\delta}(L)\mathcal{H}^{\delta}(\mathcal{B_\delta})\,,
  \end{split}
\end{equation}
where $\frac{\mathop{d\sigma_\delta}}{\mathop{d\mathcal{B_\delta}}}$ is the
fully differential Born cross section for channel $\delta$ and
$\mathcal{H}$ implements the kinematic cuts applied to the Born
phase space $\mathcal{B}$. For a 2-jet observable like thrust in DIS, there is
only one relevant partonic Born channel, corresponding to an incoming and an
outgoing quark. This also implies that the soft function $\mathcal{S}$, which
implements colour evolution, is trivial in our case.  Further, since we are
dealing with an additive observable, the multiple emission function $\mathcal{F}$
is simply given by  $\mathcal{F}(L) = e^{-\gamma_E
  R^\prime}/\Gamma(1+R^\prime)$, with $R^\prime(L)=\partial R/\partial L$ and
$R(L)=\sum_{l\in \delta} R_l(L)$.  The collinear radiators $R_l$ for the hard legs
$l$ were computed in~\cite{Banfi:2004yd} for a general observable $V$ scaling for
the emission of a soft-gluon of relative transverse momentum $k_t^{(l)}$ and
relative rapidity $\eta^{(l)}$ with respect to leg $l$ as
\begin{equation}\label{eq:CAESAR_param}
  V(k)=\left(\frac{k_{t}^{\left(l\right)}}{\mu_Q}\right)^{a}e^{-b_{l}\eta^{\left(l\right)}}d_{l}\left(\mu_Q\right)g_{l}\left(\phi\right)\,.
\end{equation}
For the case of 1-jettiness we are focusing on in this publication, we have
$a = b_l = 1$, and fixing $\mu_Q^2 = Q^2$ also $d_l g_l = 1$ since there is no
dependence on the azimuthal angle $\phi$. The precise form of the logarithm can
be varied according to
\begin{equation}\label{eq:xLscaling}
  L \to \ln\left[\frac{x_L}{v} - x_L +1\right] \to \ln \frac{x_L}{v}
  \quad\text{as}\quad v\to 0\,,
\end{equation}
to estimated the impact of sub-leading logarithms, while leaving the
distribution at the kinematic endpoint $v\sim1$ unchanged. Note this implies an
additional contribution to $R_l(L)$ to restore NLL accuracy.

The PDF factor $\mathcal{P}$, in our study applicable only to the hadronic beam,
is here given by
\begin{equation}\label{eq:pdf_ratio}
  \mathcal{P} = \frac{f_q(x,e^{-2L/(a+b)}\mu_F^2)}{f_q(x,\mu_F^2)}\,,
\end{equation}
corrects for the true initial-state collinear scale. We thereby account for the
full DGLAP evolution by calculating a simple ratio. For the purpose of matching
to a fixed order calculation, we also need the expansion of the ratio to a
given order in $\alphaS$. We generally follow the approach of
\cite{Banfi:2004yd} to implement the expansion of a leading order
approximation. This of course introduces additional effects beyond our
considered logarithmic accuracy. We argue it is safe to ignore those, given
the generally small numerical size of these contributions as seen for example in
\cite{Gerwick:2014gya}. We here for the first time apply the \Caesar implementation
in \sherpa to an observable that is sensitive to the PDF ratio (note this only applies to the
ungroomed version of thrust) and at the same time match to the {NLO
  calculation for the differential distribution and the NNLO result for the inclusive
  DIS process}. We hence need to take care of the expansion to one order
higher. Following \cite{Banfi:2004yd}, the numerator of Eq.~\eqref{eq:pdf_ratio}
can to NLL accuracy be written and expanded in powers of $\alphaS$ as
\begin{align}
  \mathbf{f}(x,e^{-2L/(a+b)}\mu_F^2) &= \exp\left[-T\left(\frac{L}{a+b}\right)\mathbf{P}\otimes\right]
  \mathbf{f} (x,\mu_F^2) \nonumber\\
  &\sim 1 - \left(T^{(1)}\left(\frac{L}{a+b}\right)+T^{(2)}\left(\frac{L}{a+b}\right)\right)
  \mathbf{P}\otimes\mathbf{f}(x,\mu_F^2) \nonumber\\
  &\phantom{= 1}+ \frac{1}{2}
  \left(T^{(1)}\left(\frac{L}{a+b}\right)\right)^2\mathbf{P}\otimes\mathbf{P}\otimes\mathbf{f}(x,\mu_F^2)
  + \mathcal{O}\left(\alphaS^3\right),\label{eq:pdf_expand}
\end{align}
where $T^{(i)}$ denotes the $i$th term obtained by expanding the integrated strong coupling
\begin{equation}
  T(L) = -\frac{1}{\pi\beta_0}\ln(1-2\alphaS\beta_0 L)
\end{equation}
in powers of $\alphaS$. The bold-faced symbols represent matrices (of splitting functions,
$\mathbf{P}$) and vectors ($\mathbf{f} = (f_u,f_d,f_s,\dots)$) in flavour space, and the
convolution is given by
\begin{equation}
  \mathbf{P}\otimes\mathbf{f} (x,\mu_F^2) = \int_{x}^1 \frac{dz}{z}
  \mathbf{P} \left(\frac{x}{z}\right) \mathbf{f} (z,\mu_F^2)\,.
\end{equation}

New terms at $\mathcal{O}(\alphaS^2)$ hence originate from the
higher order expansion of $T$, mixed terms with other parts of the resummation
multiplying the leading order expansion, and the convolution of two splitting
functions with the PDF in the last line of Eq.~\eqref{eq:pdf_expand}. The last
one is the only one that requires a non-trivial implementation. We use the
expressions from \cite{Ritzmann:2014mka} for convoluted splitting functions, and
solve the final integral for the convolution with the PDF through Monte Carlo
integration, as done at leading order.

We match our resummed calculation in the multiplicative matching scheme along
the lines of \cite{Baberuxki:2019ifp}, which we briefly recap here. The matching
to fixed order is done at the level of cumulative distributions $\Sigma(v)$. Note that we
have dropped the label for the partonic channel since in our case there is a single one
only. We expand the inclusive cross section $\sigma_\text{fo}$ as well as
the fixed-order and resummed cumulative distributions, $\Sigma_\text{fo}$ and
$\Sigma_\text{res}$ in series of $\alphaS$:
\begin{align}
  \sigma_\text{fo} & = \sigma^{(0)} + \sigma^{(1)}_\text{fo} + \sigma^{(2)}_\text{fo} + \dots\,,\\
  \Sigma_\text{fo}(v) & = \sigma^{(0)} + \Sigma^{(1)}_\text{fo}(v) + \Sigma^{(2)}_\text{fo}(v) + \dots\,,\\
  \Sigma_\text{res}(v) & = \sigma^{(0)} + \Sigma^{(1)}_\text{res}(v) + \Sigma^{(2)}_\text{res}(v) + \dots\,,
\end{align}
where the number in
parentheses indicates the respective order in $\alphaS$, and
$\sigma^{(0)}$ denotes the Born-level cross section.
Our final matched expression for the cumulative distribution, with the
dependencies on the observable value suppressed, reads:
\begin{equation}\label{eq:matching}
  \Sigma_\text{matched}=\Sigma_\text{res}
  \left(
    1+\frac{\Sigma^{(1)}_\text{fo}-\Sigma^{(1)}_\text{res}}{\sigma^{(0)}}
    +\frac{\Sigma^{(2)}_\text{fo}-\Sigma^{(2)}_\text{res}}{\sigma^{(0)}}
    -\frac{\Sigma^{(1)}_\text{res}}{\sigma^{(0)}}
     \frac{\Sigma^{(1)}_\text{fo}-\Sigma^{(1)}_\text{res}}{\sigma^{(0)}}
  \right)\,.
\end{equation}
Note that, compared to our earlier works, we use $\Sigma^{(2)}$ directly,
thus reproducing the inclusive cross section to one order
higher{, \emph{i.e.}\ NNLO}, what requires
the calculation of $\sigma^{(2)}_\text{fo}$. Importantly, the resummed \NLL
result $\Sigma_\text{res}$ is multiplied by
\begin{equation}
  \frac{\Sigma^{(1)}_\text{fo}-\Sigma^{(1)}_\text{res}}{\sigma^{(0)}} \to
  \frac{\alphaS}{2\pi}C_1 \quad\text{as}\quad v \to 0\,.
\end{equation}
{We refer to the distribution that includes all NLL terms in
  $\Sigma_\text{res}$ and additionally the coefficient $C_1$ as \NLLp accurate.}
{Through this matching procedure we achieve a formal accuracy
  of \NLOpNLLp for the differential distribution and NNLO for the
  inclusive event rate, referred to as (N)\NLOpNLLp in what follows.}

In addition to the perturbative contribution described above, there is a
significant non-perturbative component to the distribution of event shapes, that
we necessarily need to take into account in order to accurately describe
actual collider data. While it has been shown in various circumstances that soft-drop grooming
reduces the impact of hadronisation corrections, see for example
\cite{Frye:2016aiz, Baron:2018nfz, Marzani:2019evv, Baron:2020xoi,
  Caletti:2021oor, Reichelt:2021svh}, it is typically still necessary to account
for a remaining small non-perturbative contribution. We here adopt the approach of \cite{Reichelt:2021svh} to extract transfer matrices from Monte Carlo simulations. Transfer matrices are
defined as
\begin{equation}
  \mathcal{T}_{hp} = \frac{\int \mathop{dP}
    \frac{\mathop{d\sigma}}{\mathop{dP}}
    \Theta_p\left(P\right)\Theta_h\left(H(P)\right)
  }
  {\int \mathop{dP}
    \frac{\mathop{d\sigma}}{\mathop{d P}}
    \Theta_p\left(P\right)
  }\,,
\end{equation}
with
\begin{align}
  \Theta_{p}\left(P\right) &=
     \prod_{i=1}^m \theta(V_i(P)-v^\text{min}_{p,i})\theta(v^\text{max}_{p,i}-V_i(P))\,,\\
  \Theta_{h}\left(H(P)\right)&=
   \prod_{i=1}^m \theta \left( V_i\left(H(P)\right)-v^\text{min}_{h,i}\right) \theta \left(v^\text{max}_{h,i}-V_i\left(H(P)\right)\right)\,,
\end{align}
for a transition between the parton level phase space $P$
and the corresponding hadron level configuration $H(P)$,
characterised by a set of observables $V_i$ that can be calculated on both of
them. For our purpose, we assume that the requirements on the DIS kinematics,
\emph{cf.} Sec.~\ref{sec:definitions},  sufficiently fix the remaining
degrees of freedom other than 1-jettiness $\tau$. {This first of all means
  that we do not allow non-perturbative corrections to change the underlying
  Born kinematics, \emph{i.e.} the $Q^2,\,y$ bin, in contrast to, for example,
  measurements performed on jets with a potentially affected transverse momentum
  spectrum. On the other hand this implicitly assumes that it is valid to
  average the corrections for all configurations with a common 1-jettiness
  value.} Hence, we are only concerned with  events migrating between different
bins in $\tau$ within a given $Q^2,\,y$
bin. The transfer matrices as defined above can readily be extracted from the
\sherpa event generator by analysing the different stages of the events evolution,
\emph{i.e.}\ after parton showering but before hadronisation and thereafter. For practical
details of our event generation setup see Sec.~\ref{sec:sherpasetup}. Our final results
are then calculated from the resummed and matched parton level bins $\Delta\sigma_p^\text{PL}$ as
\begin{equation}\label{eq:Tfinal}
  \mathop{\Delta\sigma_h^\text{HL}} = \sum_{p} \mathcal{T}_{hp} \mathop{\Delta\sigma_p^\text{PL}}\,.
\end{equation}

\subsection{Grooming in DIS}

The framework described above has already been employed to obtain resummed predictions for
soft-drop thrust in lepton--lepton collisions at \NLOpNLLp precision~\cite{Marzani:2019evv},
for soft-drop groomed hadronic event shapes~\cite{Baron:2020xoi} and groomed jet substructure
observables at the LHC~\cite{Caletti:2021oor, Caletti:2021ysv, Reichelt:2021svh}. The
extensions made in \cite{Baron:2020xoi} to accommodate the phase space
constraints implied by soft-drop grooming, with general parameters $\zcut$ and
$\beta$, are directly applicable here. Note that \cite{Makris:2021drz} does not
define a $\beta\neq0$ version of grooming in DIS, and we make no attempt here to
extend it.

The applicability of the results from \cite{Baron:2020xoi} to DIS event shapes relies on
two statements. First, within the current hemisphere the phase space constraints to
radiation in the soft and collinear limits correspond to the case of final state radiation
in general hadronic collisions.
Second, in the beam hemisphere any soft and collinear radiation is groomed
away. Accordingly, we can treat radiation in $\mathcal{H}_B$ equivalent to the initial
state radiation case in \cite{Baron:2020xoi}, even if the precise shape of the
phase space boundary is different, but such difference does not enter at NLL
accuracy.
We analyse the behaviour of the \Centauro algorithm and the associated soft-drop
grooming variant in the language of the \Caesar framework in the following to
illustrate this.
Recall that we are working in the Breit frame. At NLL accuracy, we have to take
into account ensembles of soft particles, well separated in rapidity, around a
Born configuration consisting of the proton momentum
\begin{equation}
  P^\mu = \frac{Q}{2 x_B} n^\mu_+
\end{equation}
and the outgoing struck quark in $n_-$ direction. The virtual photon carries momentum
\begin{equation}
  q = \frac{Q}{2}(n_--n_+)~.
\end{equation}
We parameterise the momenta of additional soft gluons as
\begin{equation}
  k_i^\mu = k_t^i \left(\frac{e^{\eta_i}}{2} n_-^\mu + \frac{e^{-\eta_i}}{2} n_+^\mu + n_\perp^\mu\right)\,,
\end{equation}
where $n_\perp$ is a transverse unit vector perpendicular to $n_+$ and
$n_-$. The variable introduced in the \Centauro algorithm, \emph{cf.}
Eq.~\eqref{eq:centauro_def_zb}, can be written using the phase space variables
$\eta_i$, $k_t^i$ as
\begin{equation}
  \bar{z}_i = 2 e^{-\eta_i}~,
\end{equation}
such that the expression for the distance measure, \emph{cf.} Eq.~\eqref{eq:centauro_def}, becomes
\begin{equation}
  d_{ij} = 4 \left(e^{-2\eta_i} + e^{-2\eta_j} +
  2e^{-(\eta_i+\eta_j)}\cos\Delta\phi_{ij}\right) \sim 4
  e^{-2\eta_i}\,,
\end{equation}
where we have identified the behaviour for strong ordering in $\eta$, $\eta_i
\ll \eta_j$. In this limit, the algorithm builds up a single jet containing the
hard quark by adding the next remaining gluon that is most collinear to this
jet. The last clustering will add the gluon most collinear to the beam direction
to the jet. If all gluons are separated in rapidity well enough, there are no
other clusters to be taken care of.

From this discussion it is clear that all comparisons of scales during soft drop
will be between a soft gluon and a jet containing the hard quark. At Born level,
the four-momentum of the jet will be approximately that of the quark, and the
gluon will be the softer of the two. With this in mind the hardness measure for
soft drop for soft momentum $k_i$ can be written as
\begin{equation}
  z_i \sim \frac{k_t^i}{Q}e^{\eta_i}~.
\end{equation}
Within the current hemisphere,
the phase space restriction, on an emission that passes the soft-drop criterion,
is given by
\begin{equation}\label{eq:sd_cond_soft}
  \frac{k_t e^\eta}{Q} > \zcut~,
\end{equation}
which precisely matches the one given in \cite{Baron:2020xoi}  for $\beta = 0$
(see Sec. 3.4 point (iv), and note that the hard quark has energy $Q/2$ in the
Breit frame).

Note that particles outside of the current hemisphere will enter in
Eq.~\eqref{eq:sd_cond_soft} with negative rapidity $\eta$. They will hence be
groomed away unless they are at very high $k_t$, only causing logarithms of
$\zcut$. We note again that the precise shape of the phase space boundary is
different from what is given in \cite{Baron:2020xoi} for initial states. The
main point is however that only logarithms of $\zcut$ are produced, which we
ignore noting again that we work in the limit $v \ll \zcut$.

\subsection{Calculational tools and setup}

As already stated, the resummation calculation for 1-jettiness is accomplished
with the \Caesar plugin to \sherpa that hooks into the event generation
framework\footnote{Note, during the course of this work the plugin has been
ported to the \sherpa-3.0 release series.}. \sherpa thereby provides all the process
management, and gives access to the \comix matrix element generator~\cite{Gleisberg:2008fv},
as well as phase-space integration and event-analysis functionalities. We make use of \sherpa's
interface to \LHAPDF \cite{Buckley:2014ana} and use the PDF4LHC\_40\_pdfas PDF
set, as we do for the parton-shower simulations outlined in the previous section.
The value of the strong coupling is set accordingly, \emph{i.e.}\
$\alpha_S(M^2_Z)=0.118$. The \sherpa framework is also used to compile all the required
higher-order tree-level and one-loop calculations. For the NLO QCD computations we use the
\sherpa implementation of the Catani--Seymour dipole subtraction~\cite{Gleisberg:2007md}
and the interfaces to the \recola~\cite{Actis:2016mpe,Biedermann:2017yoi} and
\OpenLoops~\cite{Cascioli:2011va} one-loop amplitude generators. The calculation
of NNLO accurate predictions for DIS has been automated in \sherpa in
\cite{Hoche:2018gti}, and we use it to compute cross sections $\sigma^{(2)}_{\text{fo}}$
at order $\alphaS^2$ differential in $Q^2$ and $y$ to achieve
overall NNLO accuracy for inclusive cross sections.
This corresponds to an accuracy of the distribution differential in thrust at \NLO,
and we refer to the combined accuracy of our fixed order predictions including
cross sections as (N)\NLO. The plugin implements the building blocks of the \Caesar
master formula Eq.~\eqref{eq:CAESAR}, along with the necessary expansion in $\alpha_s$
used in the matching with fixed-order calculations. The building blocks are evaluated
fully differentially for each Born-level configuration $\mathcal{B}_\delta$ of a
given momentum configuration. Jet clustering and grooming functionalities are
accessed through the  interface of \sherpa to \fastjet~\cite{Cacciari:2011ma}.
Non-perturbative corrections are extracted from dedicated runs of the \sherpa
generator using the identical setup described in Sec.~\ref{sec:sherpasetup},
thereby employing the functionality of the \rivet analysis tool to provide
access to intermediate evolution stages through the \HepMC event record~\cite{Dobbs:2001ck}.

\section{Results for (groomed) 1-jettiness in DIS}\label{sec:results}

Having outlined our calculational techniques for describing hadronic
final state observables in neutral current DIS, we can finally present our numerical results
for the 1-jettiness event shape. We begin by discussing selected
results for the ungroomed case. We have compiled predictions for a wide range of
$Q^2$ values, \emph{i.e.}\ $Q^2\in[150,20000]\;\text{GeV}^2$. Furthermore, we consider
  the production cross section differential in the events inelasticity, thereby covering
  the region $y\in[0.05,0.94]$. For brevity, we here focus on three kinematic regions
  corresponding to medium values of $y\in [0.4,0.7]$ and rather low ($Q^2\in [150,200]\,\text{GeV}^2$),
  medium ($Q^2\in [440,700]\,\text{GeV}^2$) and high ($Q^2\in[3500,8000]\,\text{GeV}^2$) photon
  virtuality.

  Along with the central predictions we show error bands indicating the perturbative uncertainty obtained from
7-point variations of $\mu_R, \mu_F$, in both the shower and the semi-analytic
calculation, and in addition a variation of $x_L = 0.5,2$ in the latter,
\emph{cf.}~Eq.~\eqref{eq:xLscaling}.
Furthermore, we include an uncertainty estimate related to the tuning of beam-fragmentation
parameters based on replica tunes, see Sec.~\ref{sec:tuning}. Generally, this
contribution is found to be rather small compared to the perturbative
uncertainties. We observe the overall uncertainties for the NLO QCD
matrix element plus parton-shower simulations and the resummation predictions
to be of similar sizes.

\begin{figure}
  \includegraphics[width=.32\textwidth]{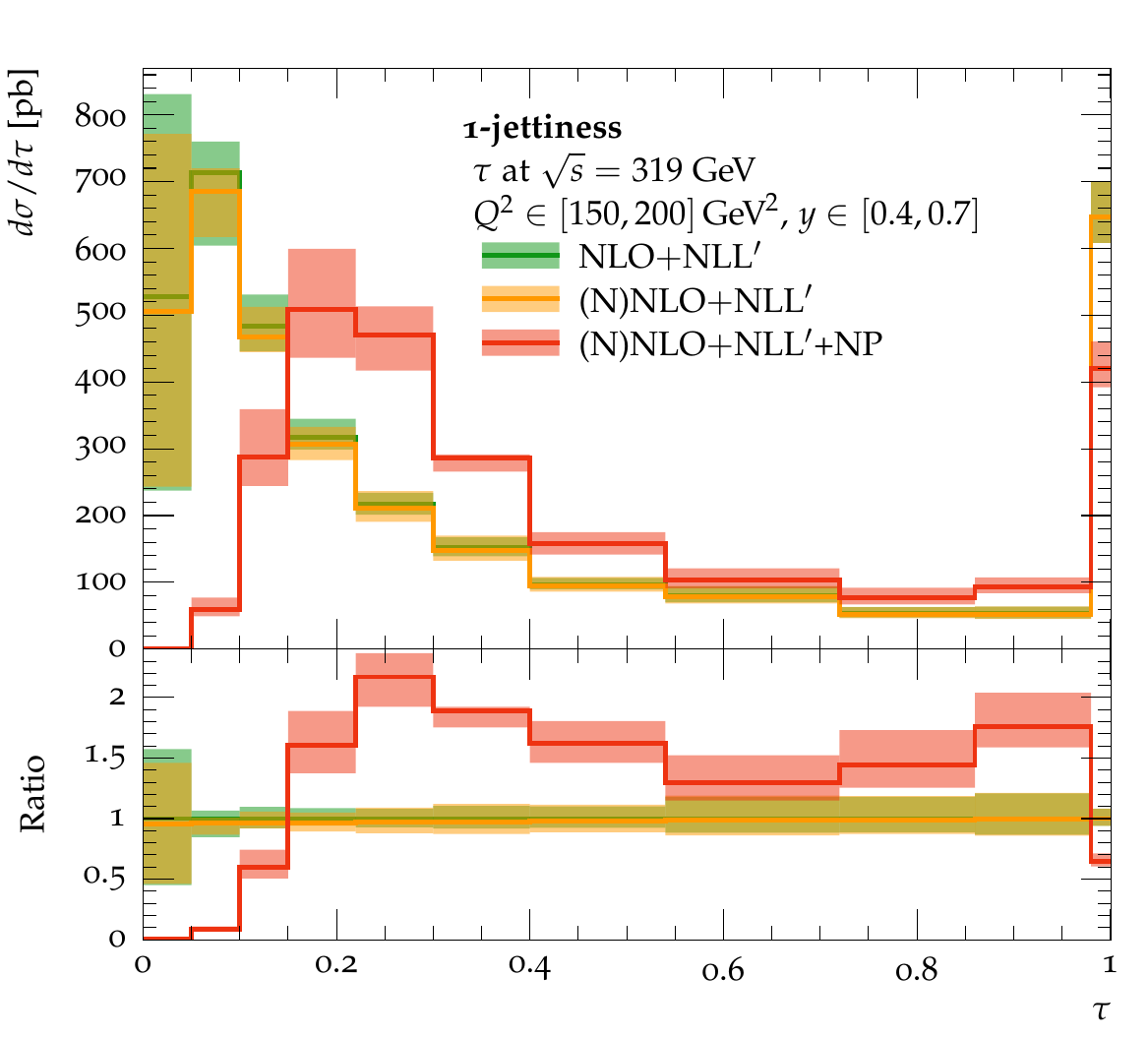}
  \includegraphics[width=.32\textwidth]{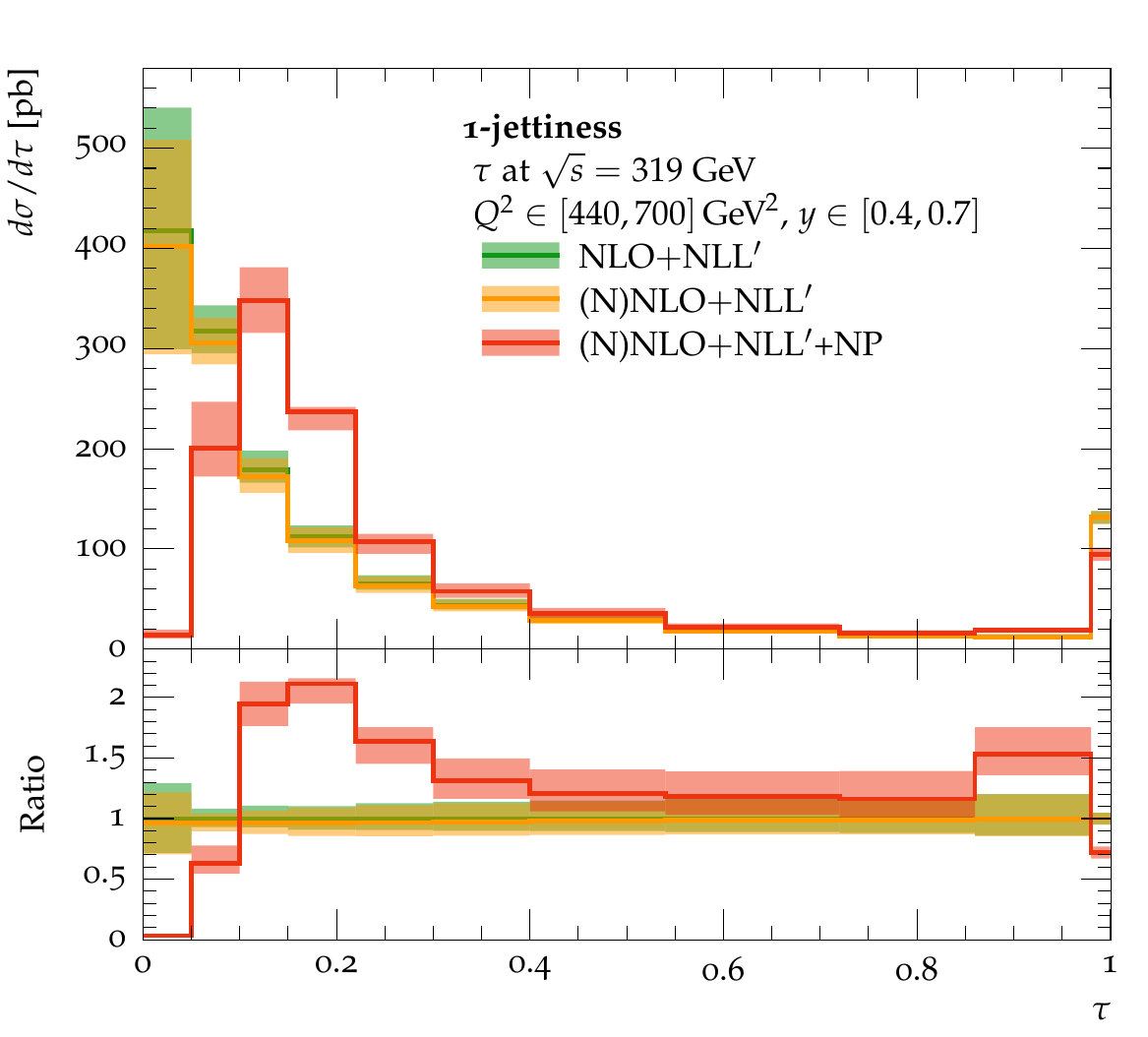}
  \includegraphics[width=.32\textwidth]{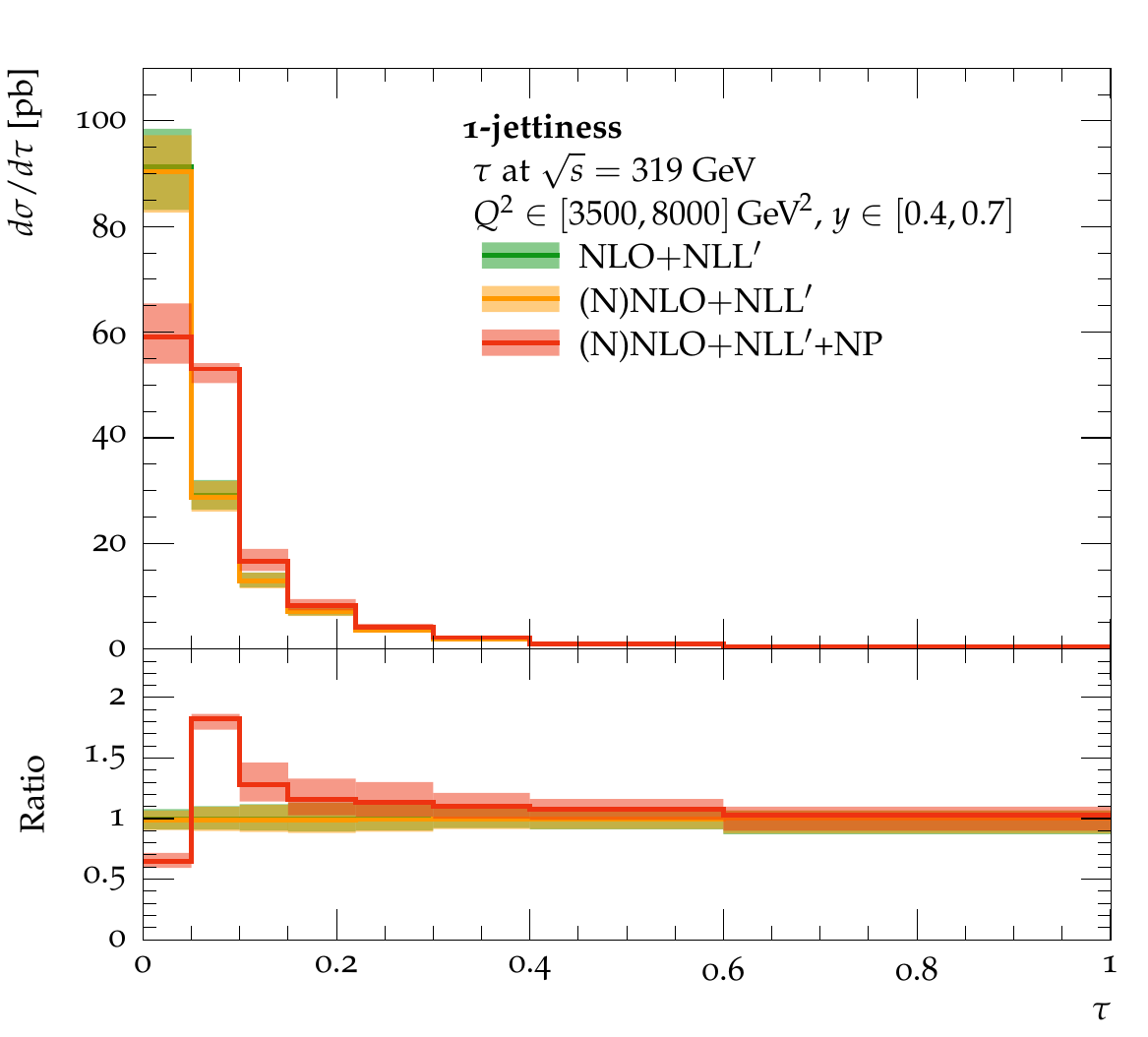}
  \caption{Distributions of ungroomed 1-jettiness in selected $y-Q^2$ bins, at different
    stages of the calculation, at \NLOpNLLp accuracy, including the
    normalisation at NNLO ((N)\NLOpNLLp) accuracy, and including
    non-perturbative corrections. All results correspond to DIS kinematics with
    $y\in [0.4,0.7]$ and the plots represent from left to right regions of $Q^2/\text{GeV}^2\in[150,200]$, $[440,
      700]$, and $[3500,8000]$, respectively. The lower panels present the ratio
    to the plain \NLOpNLLp result.}\label{fig:res_stages}
\end{figure}

We first analyse the behaviour of the \NLOpNLLp resummation calculation upon inclusion of
the NNLO normalisation correction and non-perturbative effects. To this end we compile
in Fig.~\ref{fig:res_stages} corresponding predictions for the three kinematic regions
specified before. From the lower panels, showing the ratio to the respective \NLOpNLLp
result, it can be read off, that correcting the normalisation to NNLO accuracy has a rather
small impact. The differential cross section receives a small negative correction, of at
most a few percent at small $\tau$ in the lower $Q^2$ region. Note, however, that even the
smallest $Q^2$ values in this analysis remain sizeable compared to the overall range accessible
for the HERA experiments. Somewhat more significant is the reduction in the perturbative
uncertainties when going from NLO to NNLO, in particular for the bulk of the distributions,
\emph{i.e.}\ low values of 1-jettiness.

Next, we consider the inclusion of non-perturbative corrections based on the transfer-matrix
approach described in Sec.~\ref{sec:res_review}. As clearly visible in Fig.~\ref{fig:res_stages}
these significantly alter the shape of the distributions, introducing a sizeable shift towards
larger 1-jettiness values. In particular for the low and medium $Q^2$ region the first bin
gets almost entirely depopulated. In contrast, for values of $\tau \approx 0.1\dots 0.2$
corrections can reach up to $+100\%$. The effect of hadronisation corrections is
less pronounced at higher $Q^2$. We furthermore note, that the non-perturbative corrections
through the bin migration via transfer matrices partially compensate the dependence of the
perturbative calculation on scale variations and in particular of $\mu_R$.

We close this first discussion of the resummed predictions for ungroomed 1-jettiness
by pointing to the distinct peak at $\tau\approx 1$ for the low and medium $Q^2$
distributions, emerging after a significant decline of the differential cross section
from lower to larger observable values. For the given observable definition the
configuration $\tau=1$ can be attributed to events with an empty current hemisphere
${\mathcal{H}}_C$~\cite{Kang:2014qba}. Such configurations first appear when considering
the NLO real-emission correction to the DIS process, when both final state partons feature
negative longitudinal momenta in the Breit frame, such that 1-jettiness defaults to 1,
see Eq.~\eqref{eq:tau_def}. We here account for these configurations through matching to
the exact NLO QCD result for $\tau$, \emph{i.e.}\ including the full ${\cal{O}}(\alpha_S)$
corrections to the two-parton channel. It can be observed, that hadronisation corrections
reduce the amount of $\tau\approx 1$ events, what can be expected, as the fragmentation of partons
originally in the beam hemisphere might spill over hadrons in the current hemisphere.

\begin{figure}
  \includegraphics[width=.32\textwidth]{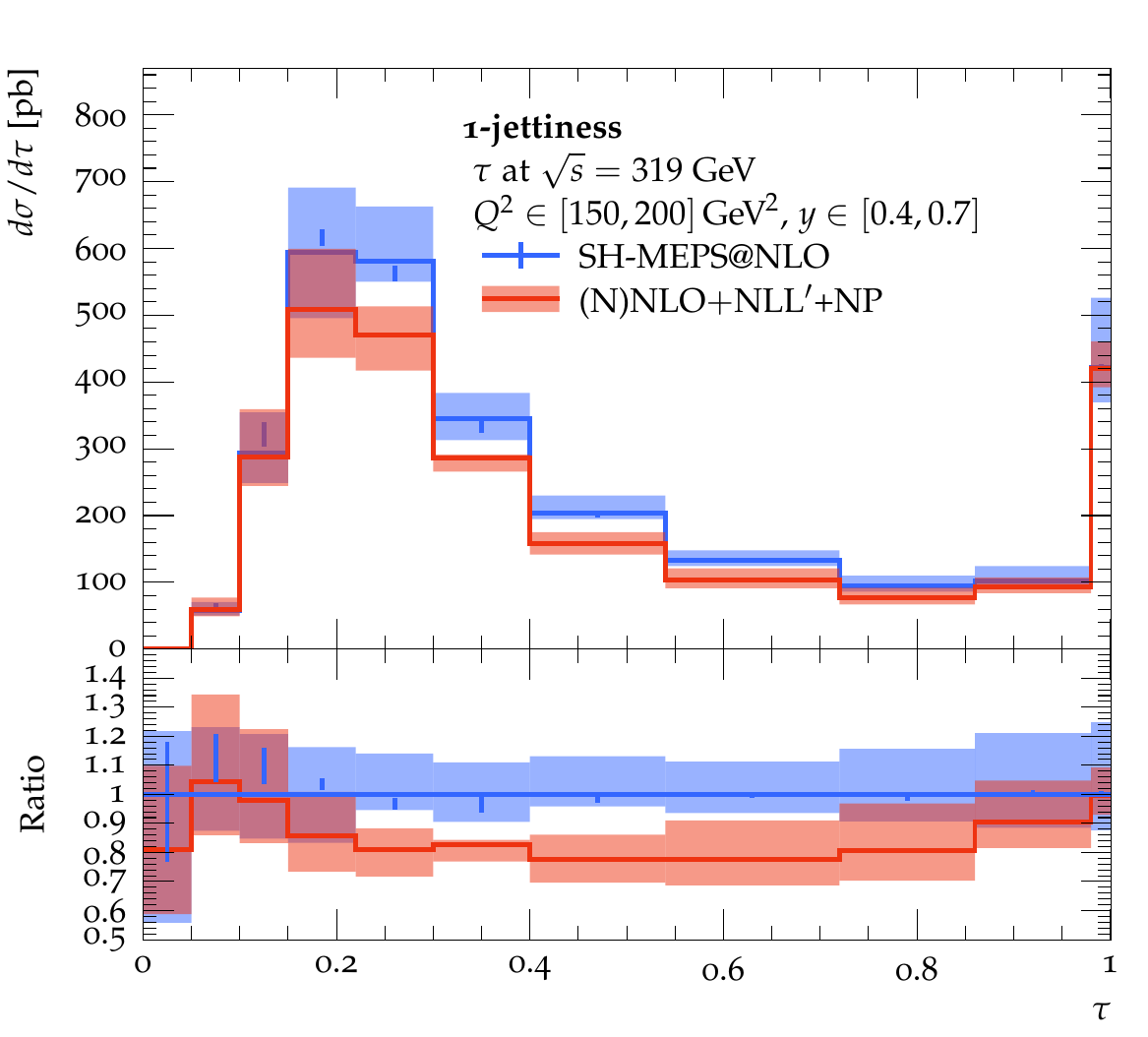}
  \includegraphics[width=.32\textwidth]{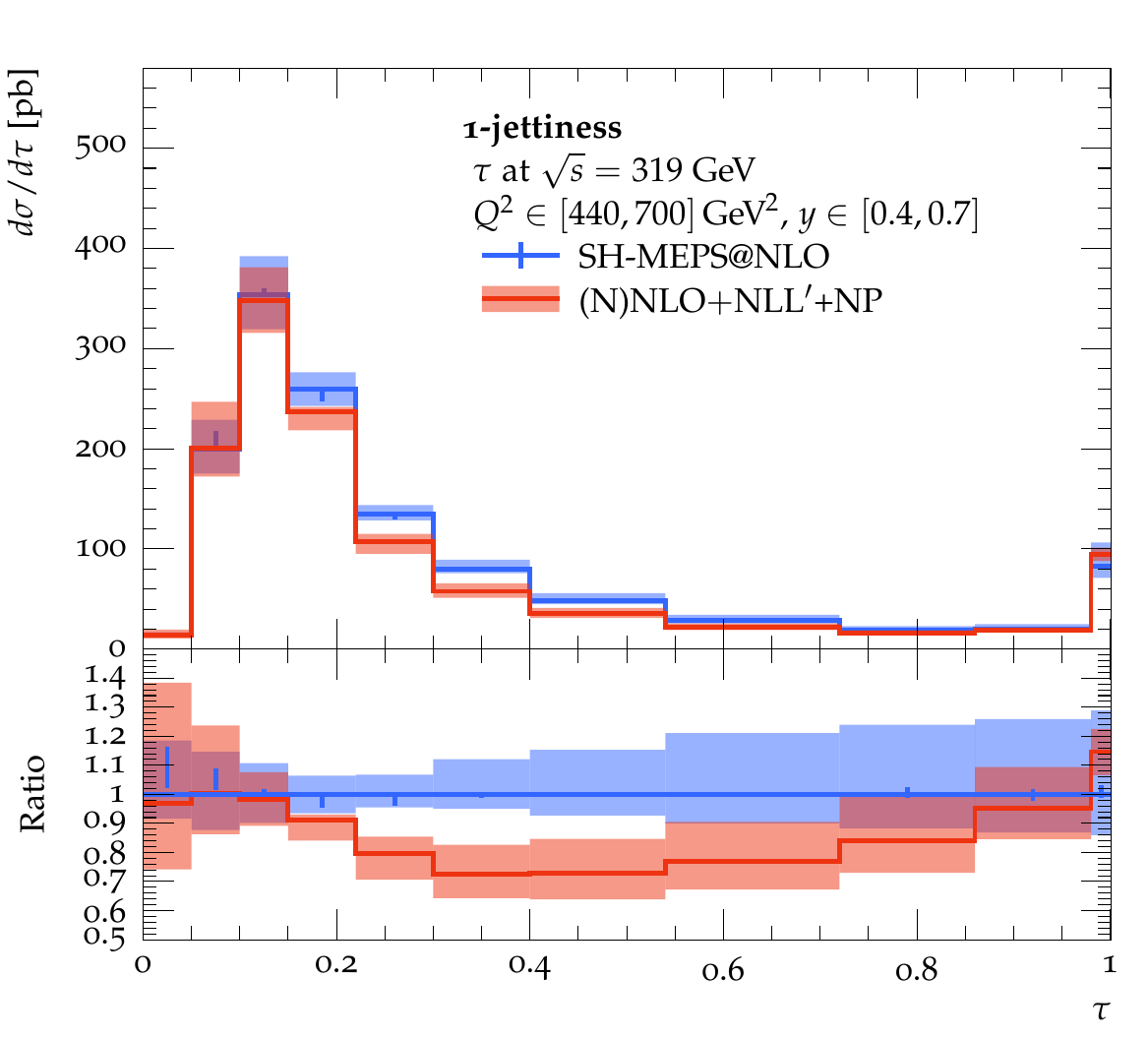}
  \includegraphics[width=.32\textwidth]{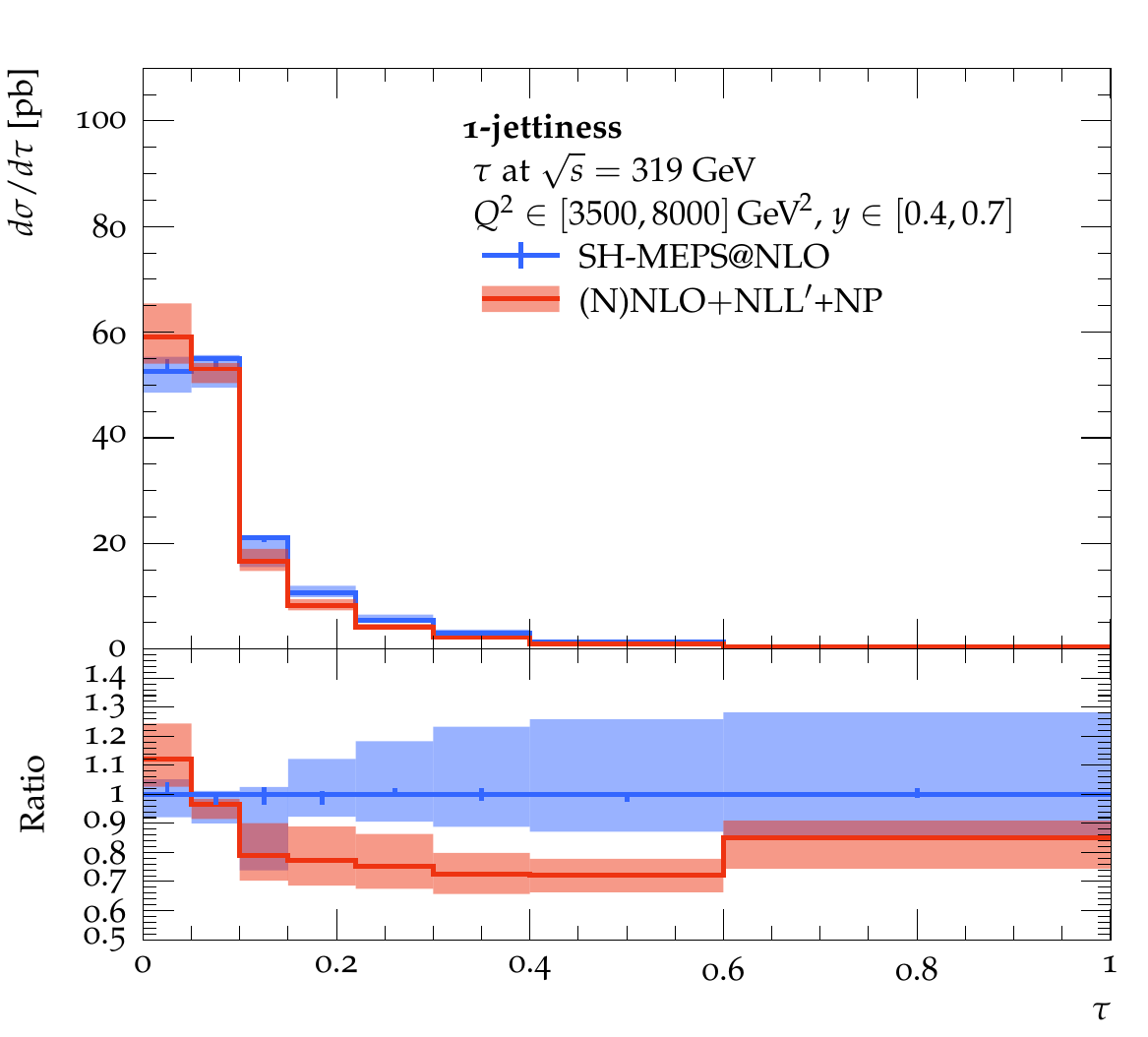}
  \caption{Distributions of 1-jettiness in selected $y-Q^2$ bins, \emph{i.e.}\
    $y\in [0.4,0.7]$ and, from left to right, $Q^2/\text{GeV}^2\in[150,200]$, $[440,
      700]$, and $[3500,8000]$, respectively. Shown are hadron level MEPS@NLO predictions
    from \sherpa and results at (N)\NLOpNLLpNP accuracy. The lower panels present the ratio
    to the MEPS@NLO result.}\label{fig:res_nogroom}
\end{figure}

We now turn to the presentation of the hadron level results from MEPS@NLO simulations
with \sherpa as outlined in Sec.~\ref{sec:sherpasetup} and compare those to the
(N)\NLOpNLLpNP predictions. In Fig.~\ref{fig:res_nogroom} we compare the respective
results for the three considered kinematic regions. We observe an overall fair agreement
between the matrix element improved shower simulations at hadron level obtained from
\sherpa and the resummed and matched calculation at (N)\NLOpNLLpNP accuracy, corrected for
non-perturbative effects. In general the merged prediction features a somewhat harder
spectrum, \emph{i.e.}\ favours somewhat larger observable values. This might also be
attributed to the inclusion of the exact tree-level three- and four-jet matrix elements,
see Eq.~\eqref{eq:MEPS_MEs}. These contributions feature LO scale dependence and are thus
the source for the somewhat enlarged theoretical uncertainties in the shower simulation towards
larger values of $\tau$. However, the regions of small 1-jettiness agree within uncertainties
for all three kinematic regions, up until the peak of the respective distribution. Towards the
kinematic endpoint, the two approaches tend to agree again, with both calculations predicting
very similar cross sections for events with $\tau\sim1$.

\begin{figure}
  \includegraphics[width=.32\textwidth]{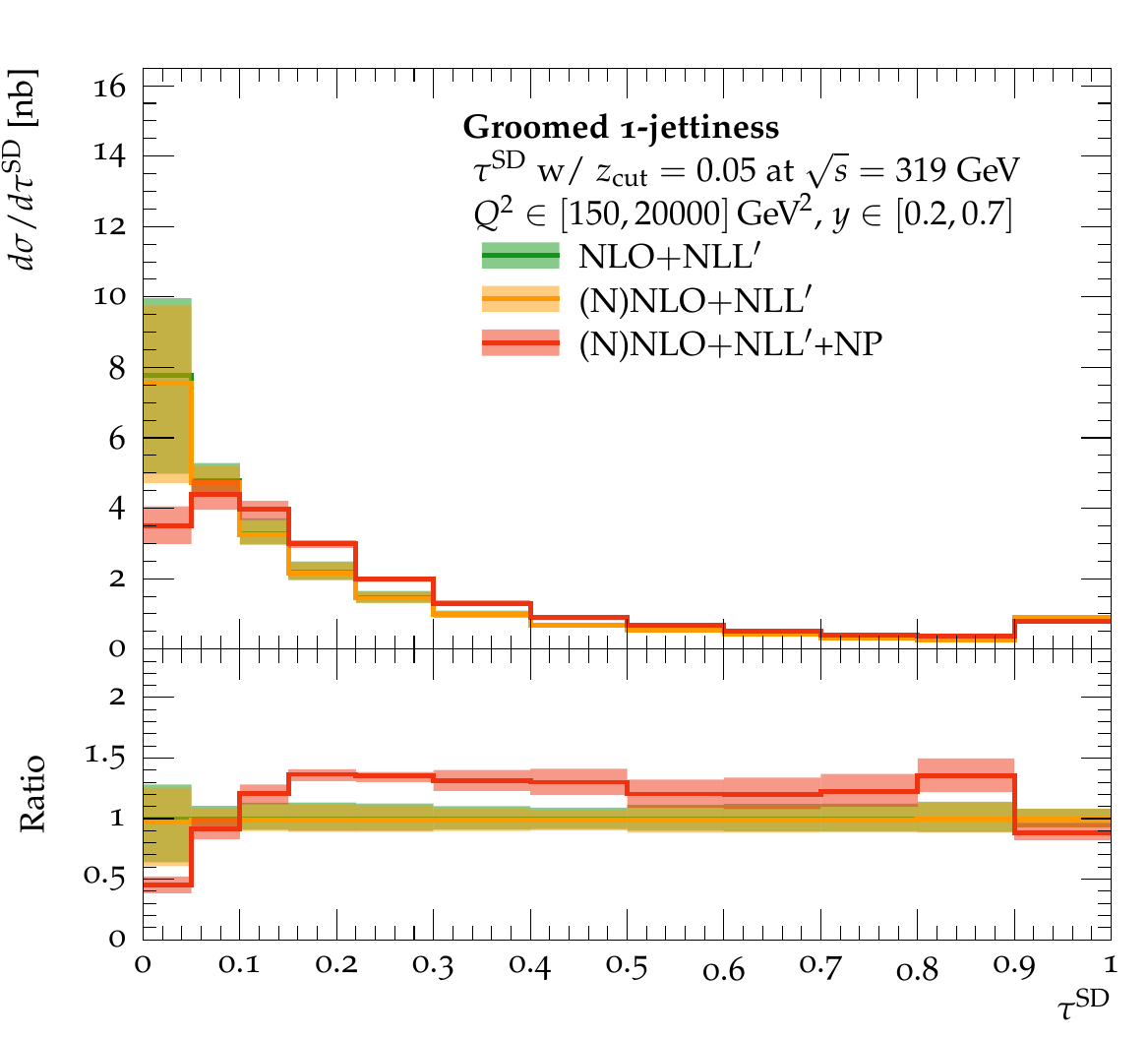}
  \includegraphics[width=.32\textwidth]{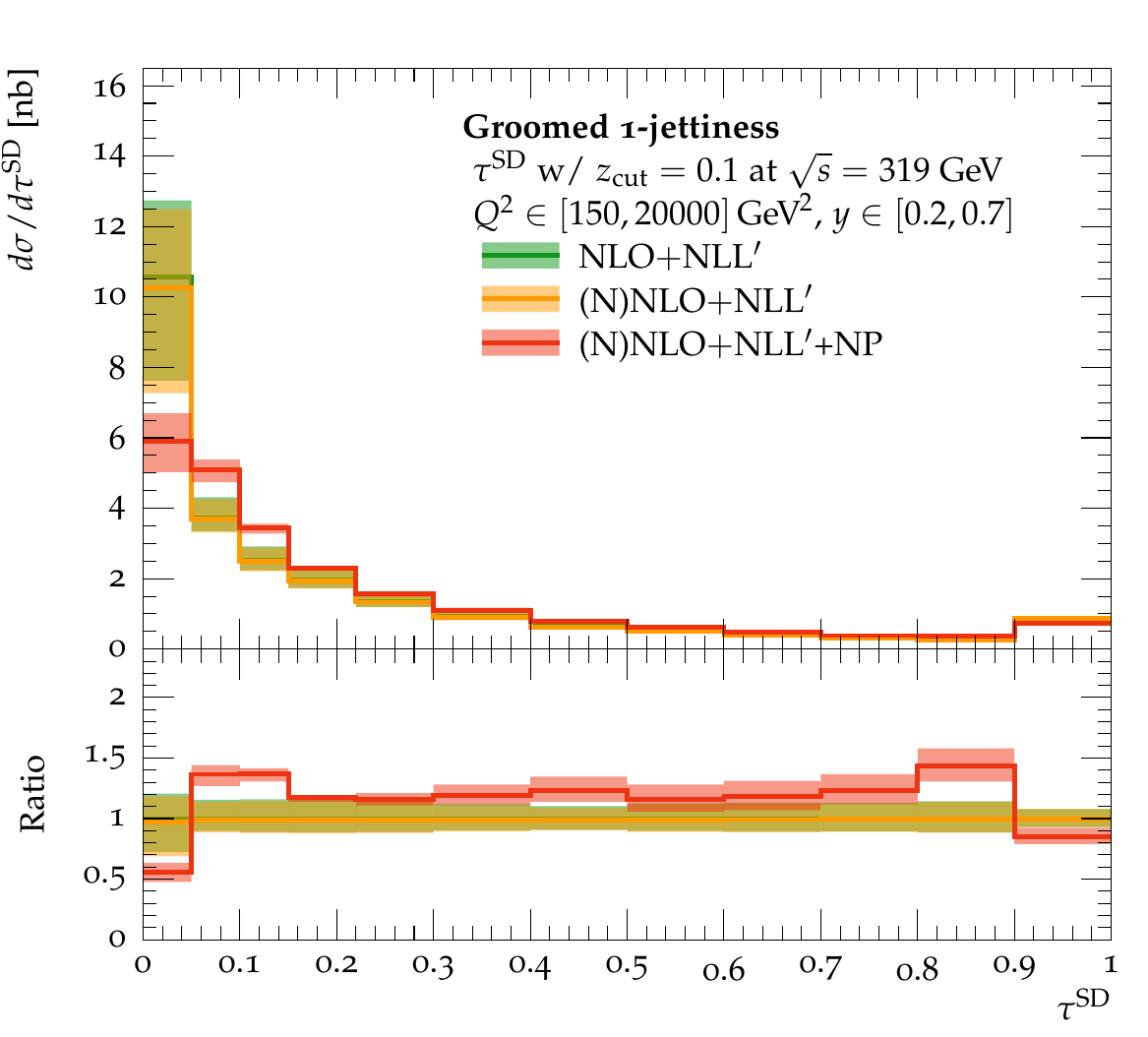}
  \includegraphics[width=.32\textwidth]{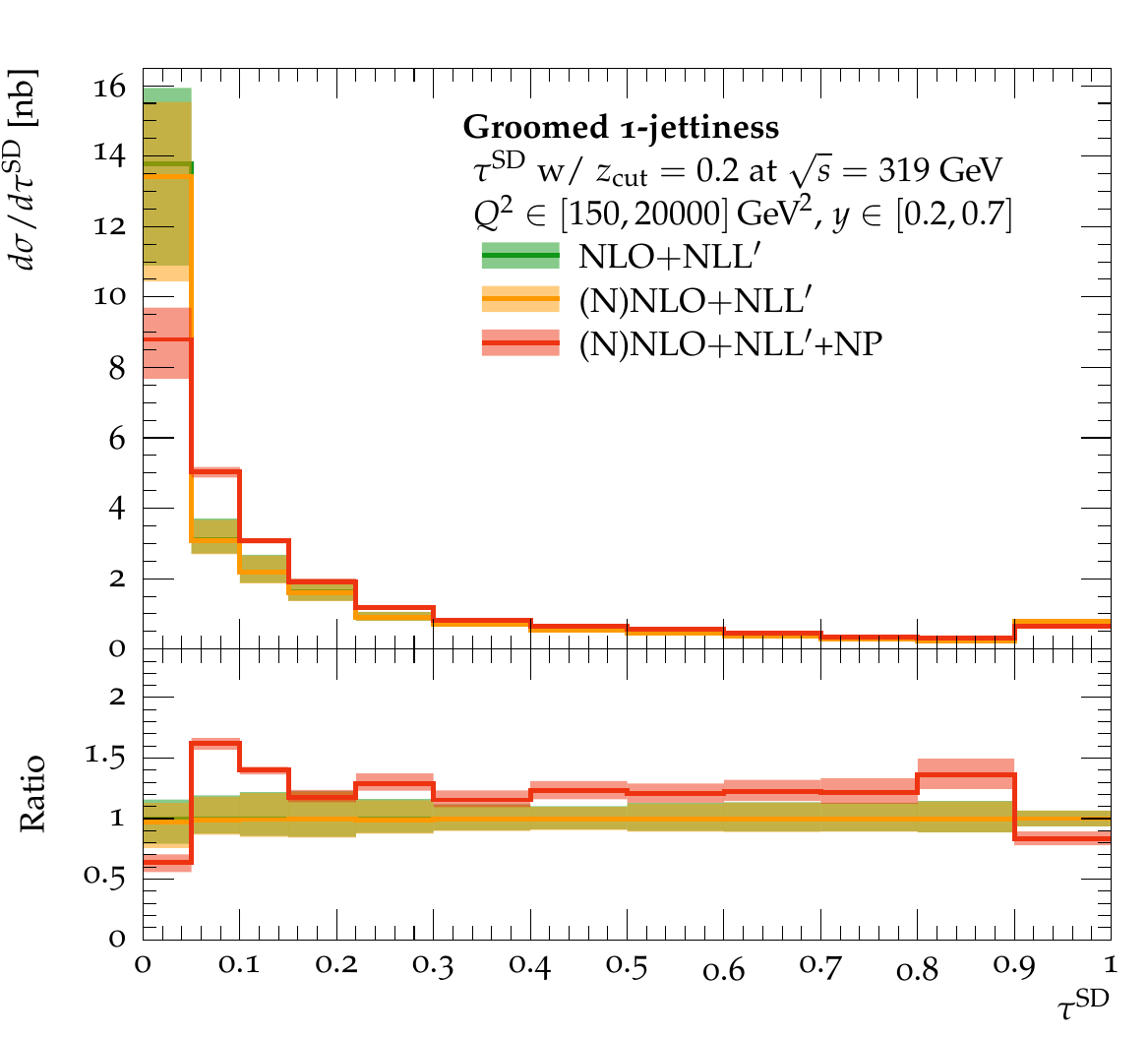}
  \caption{Distributions of groomed 1-jettiness, at different
    stages of the calculation, at \NLOpNLLp accuracy, including the
    normalisation at NNLO ((N)\NLOpNLLp) accuracy, and including
    non-perturbative corrections. From left to right the plots
    represent predictions for the grooming parameter $\zcut = 0.05,0.1,0.2$,
    respectively. The lower panels present the ratio
    to the plain \NLOpNLLp result.}\label{fig:res_groom_stages}
\end{figure}

Besides the plain 1-jettiness event shape we here also consider the effect of
soft-drop grooming the hadronic final state. In Fig.~\ref{fig:res_groom_stages}
we show resummed predictions for groomed 1-jettiness, referred to as $\tau^{\text{SD}}$
in what follows, integrated over the full $Q^2$ range,
\emph{i.e.}\ $Q^2\in[150,20000]\;\text{GeV}^2$, and the inelasticity region
  $y\in[0.2,0.7]$. We compiled predictions for three commonly considered values of $\zcut$,
  namely $\zcut=0.05,0.1,0.2$, thereby always assuming the angular grooming parameter
  $\beta=0$. As seen for the ungroomed case, we note rather small effects of the NNLO
  normalisation corrections compared to the \NLOpNLLp calculation. Also the systematic
  uncertainties hardly change from NLO to NNLO. However, the size of the non-perturbative
  corrections is significantly reduced relative to the ungroomed case, staying below $50\%$
  and being largely flat over a wide range of $\tau^\text{SD}$, apart from very low values of
  1-jettiness and at the endpoint $\tau^\text{SD}\sim 1$. This confirms the potential of
  soft-drop grooming to mitigate hadronisation effects for event shape observables also
  in DIS, seen before in $e^+e^-$~\cite{Baron:2018nfz,Marzani:2019evv}
  and $pp$ collisions~\cite{Baron:2020xoi}.
  
\begin{figure}
  \includegraphics[width=.32\textwidth]{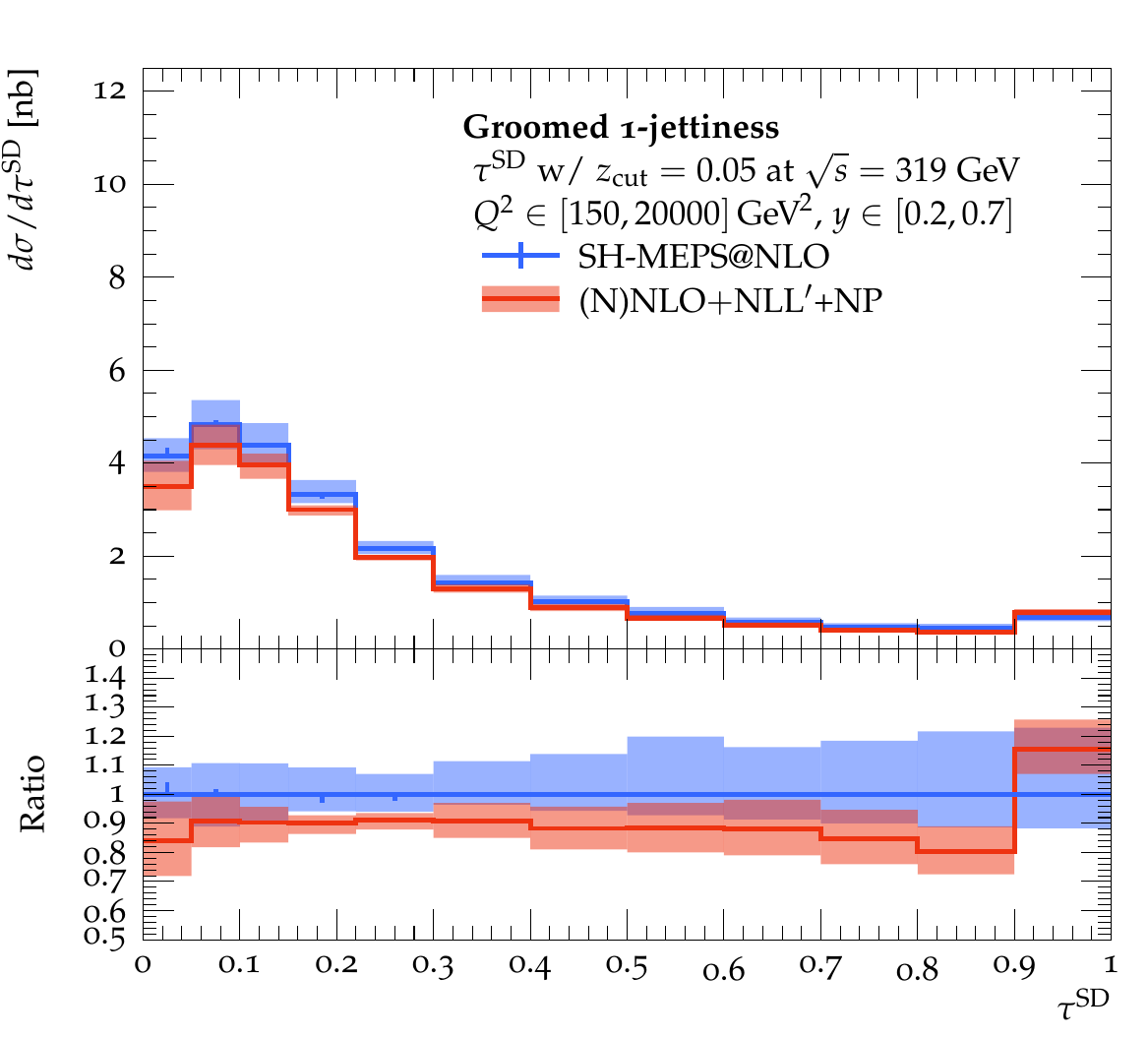}
  \includegraphics[width=.32\textwidth]{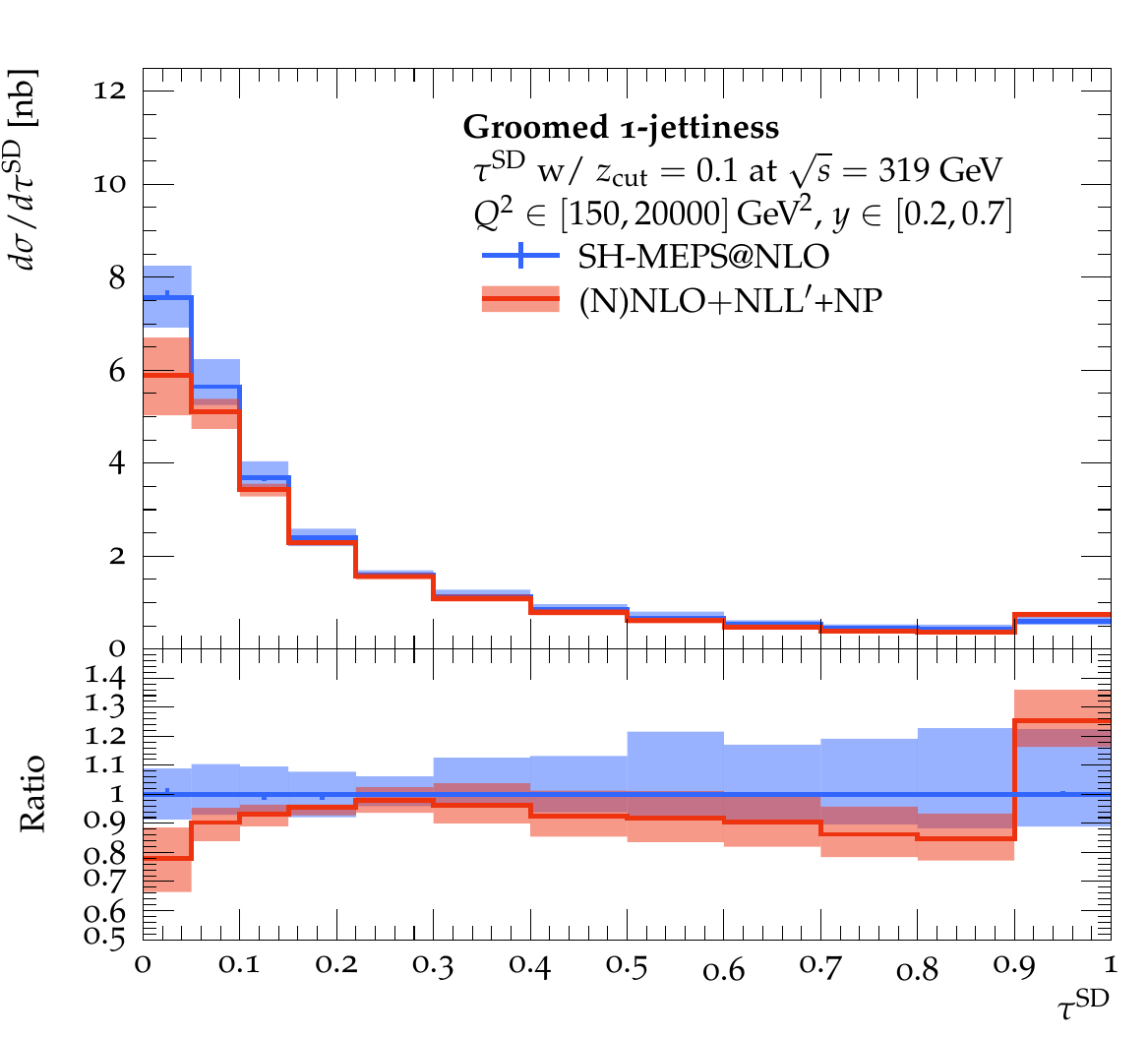}
  \includegraphics[width=.32\textwidth]{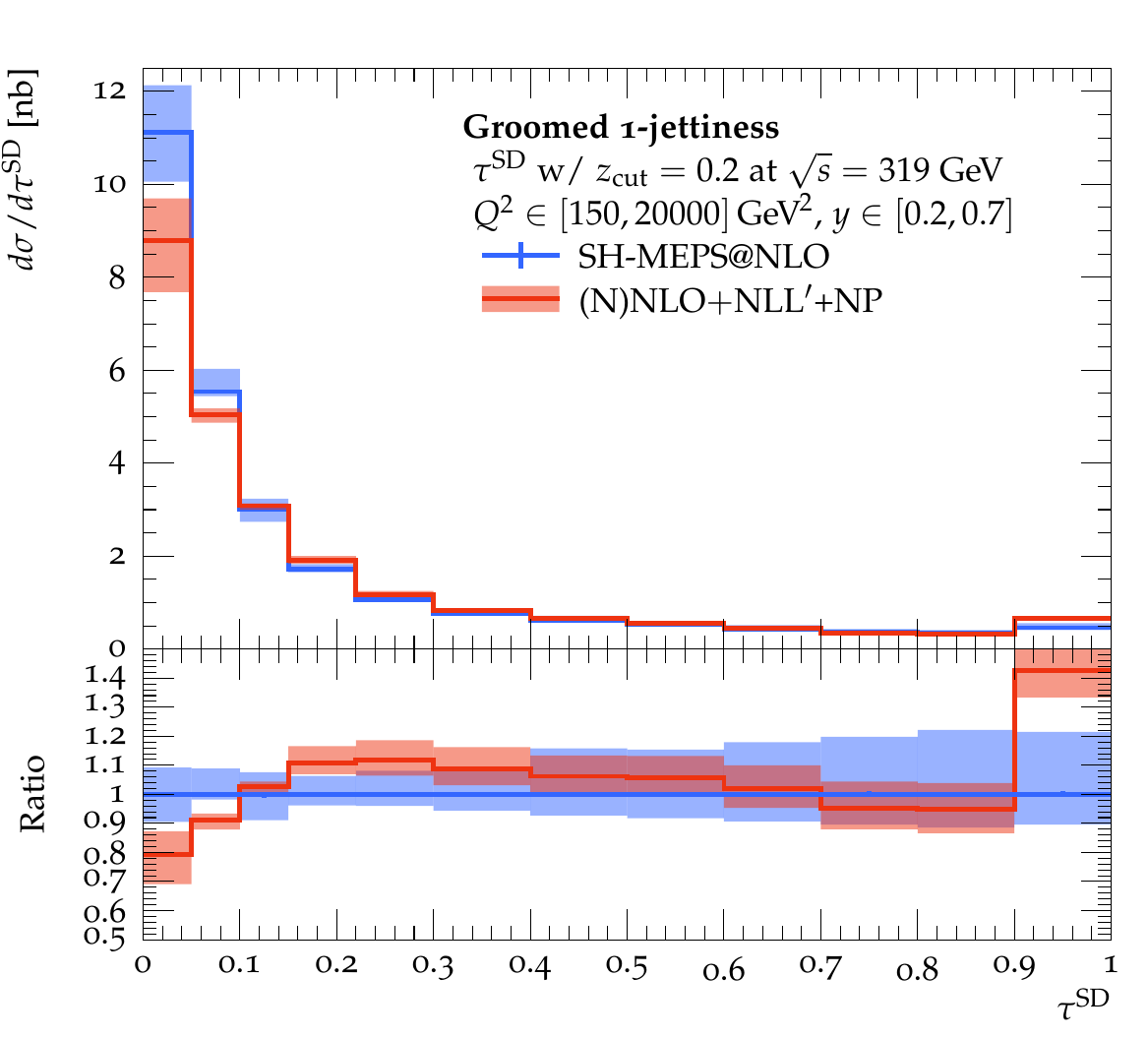}
  \caption{Distributions of groomed 1-jettiness. Shown are hadron level MEPS@NLO predictions
    from \sherpa and results at (N)\NLOpNLLpNP accuracy. From left to right the plots
    represent predictions for the grooming parameter $\zcut = 0.05,0.1,0.2$,
    respectively. The lower panels present the ratio to the MEPS@NLO result.}\label{fig:res_groom}
\end{figure}

The comparison of the (N)\NLOpNLLpNP results with hadron level simulations at MEPS@NLO
accuracy is presented in Fig.~\ref{fig:res_groom}. For all the $\zcut$ values, we observe
good agreement between our \sherpa simulation and the resummation calculation somewhat better
than for the ungroomed case. In all three cases, the (N)\NLOpNLLpNP calculation
predicts a larger cross section in the $\tau\sim1$ bin, although still
compatible within the uncertainty of the event generator for $\zcut=0.05$ and
the combined uncertainty for both calculations for $\zcut=0.1$. Apart from this
last bin, for these two $\zcut$ values the resummation calculation is consistently
below the \sherpa simulation. In the case of $\zcut=0.05$, this happens flat over the
full spectrum $\tau^\mathrm{SD}<1$, while for increasing $\zcut$ a slight shape develops,
with the (N)\NLOpNLLpNP cross section decreasing faster for $\tau^\mathrm{SD}<\zcut$
than what is seen in the Monte Carlo simulation.

It will be interesting to compare the (N)\NLOpNLLpNP predictions and the \sherpa MEPS@NLO
simulations with the data of upcoming measurements by the H1 experiment. This will shed
light on the found deviations between the two sets of predictions and possibly guide the
development of yet improved theoretical predictions, \emph{e.g.}\ through the inclusion
of next-to-next-to-leading logarithmic corrections.

\section{Conclusions}\label{sec:conclusions}

We presented the calculation of theoretical predictions for the 1-jettiness
event shape in neutral current DIS at HERA energies. The here considered 1-jettiness observable,
evaluated in the Breit frame, is equivalent to the well-known thrust variable that has been widely
studied at lepton and hadron colliders. Besides plain 1-jettiness we also considered its
variant after soft-drop grooming the hadronic final state using different values of the
grooming parameter $\zcut$. We consider the triple-differential cross section in the observable,
momentum transfer $Q^2$, and the events inelasticity $y$.

Based on the \Caesar formalism we derive NLL accurate results matched to the exact NLO QCD matrix
element for the two-jet DIS matrix element. Furthermore, we include the exact NNLO QCD corrections
to the inclusive DIS process, thereby achieving full NNLO accuracy for the integrated observable
distribution. We furthermore correct our results of (N)\NLOpNLLp accuracy for non-perturbative
hadronisation effects through a transfer matrix that takes into account migration in the observable
value when going from parton to hadron level. The corresponding corrections have been extracted
from Monte Carlo simulations at MEPS@NLO accuracy with the \sherpa generator. To this end, we have
performed tunes of the beam-fragmentation parameters of \sherpa's new cluster fragmentation model
against data from the H1 and ZEUS experiments. We thereby also derived replica tunes that account
for the parametric uncertainties.

For plain 1-jettiness we have shown results for three kinematic regions, corresponding to medium
inelasticity $y$ and ranges of rather low, medium, and high $Q^2$ values. While the impact of the NNLO
contributions is found to be very small, hadronisation corrections significantly sculpt the
differential distributions, pushing events from lower to larger 1-jettiness values. When comparing
the hadronisation corrected (N)\NLOpNLLp predictions with hadron level predictions from \sherpa
good agreement is found, with larger deviations dominantly in the region $0.2<\tau<0.6$.
Quite good agreement is found regarding events at the endpoint of the distribution, \emph{i.e.}\
$\tau\simeq 1$. For the low and medium $Q^2$ regions the distribution here develops a significant
peak, that can be attributed to events with an empty current hemisphere.

For the soft-drop groomed variant of 1-jettiness we have shown predictions for three values of
$\zcut$, integrated over a wide range of $Q^2$, \emph{i.e.}\  $Q^2\in[150,20000]\;\text{GeV}^2$,
and $y\in[0.2,0.7]$. For all values of $\zcut$ non-perturbative corrections to the
resummed predictions get significantly reduced, when comparing to the ungroomed case.
Furthermore, an improved agreement with the hadron level predictions from \sherpa is found.

{It will be exciting to confront the two types of predictions with actual data from the
HERA collider that are currently being analysed by the H1 experiment. We can expect that in particular
for the ungroomed 1-jettiness observable data should be able to discriminate between the two predictions.
This will motivate and guide the development and advancement of the theoretical predictions.
For DIS parton shower simulations there are recent developments
towards the inclusion of NNLO QCD corrections~\cite{Hoche:2018gti} and to achieve formal NLL
accuracy~\cite{Dasgupta:2020fwr, Forshaw:2020wrq, Herren:2022jej, vanBeekveld:2023lfu}.
This would allow to match the precision of the analytic predictions we presented in this
study. Improving the analytic calculation might require the inclusion of higher-logarithmic corrections
or improved means to account for non-perturbative corrections. Furthermore, a detailed analysis of
systematic differences between analytic NLL resummation and shower algorithms implementing unitarity
and momentum conservation along the lines of \cite{Hoche:2017kst} might help to pin down the origin
of the observed differences.}

\section*{Acknowledgements}
We would like to thank Daniel Britzger and Henry Klest for triggering us to dive into DIS event shapes
and a very fruitful communication. We furthermore thank Johannes Hessler and Vinicius Mikuni for discussions.
We are indebted to Stefan H\"oche for assistance with the NNLO corrections and we are grateful to Frank Krauss for
help with \sherpa's new beam fragmentation model. \\

MK and SS acknowledge support from BMBF (05H21MGCAB) and funding by the Deutsche Forschungsgemeinschaft
(DFG, German Research Foundation) - project number 456104544 and 510810461.
DR is supported by the STFC IPPP grant (ST/T001011/1).


\appendix
\section{Tuning details}
\label{app:tuning-details}

We here collate more detailed information on the tuning of the \ahadic beam-fragmentation
parameters. The \rivet analyses and considered observable measurements by the H1 and ZEUS
HERA experiments used for the tuning are summarised in Tab.~\ref{tab:tuning_analysis}.

\begin{table}
  \centering
  \begin{tabular}{l l l}
    \toprule
    \rivet Analysis name [reference] & Observables & {Virtuality range [$\text{GeV}^2$]} \\
    \midrule
        {\tt H1\_2006\_I699835 } \cite{H1:2005zsk}
        & thrust, jet broadening
        & {$Q^2\in [256,400]$} \\

        {\tt H1\_1994\_S2919893} \cite{H1:1994vjx}
        & transverse energy flow
        &{$Q^2\in[10,100]$}\\

        & energy-energy correlation
        &{$Q^2\in[10,100]$}\\

        {\tt H1\_1995\_I394793 } \cite{H1:1995cqf}
        & quark fragmentation functions
        & {$Q^2>100$} \\

        {\tt H1\_1996\_I422230 } \cite{H1:1996ovs}
        & charged multiplicity distributions
        & {$Q^2\in[10,1000]$} \\

        {\tt H1\_1996\_I424463 } \cite{H1:1996muf}
        & charged particle spectra
        & {$Q^2\in[5,50]$} \\

        {\tt H1\_1997\_I445116 } \cite{H1:1997mpq}
        & quark fragmentation functions
        & {$Q^2\in[100,8000]$}\\

        & charged hadron energy spectra
        & {$Q^2\in[100,8000]$}\\

        {\tt H1\_2000\_S4129130} \cite{H1:1999dbb}
        & transverse energy flow
        & {$Q^2\in[10,2200]$} \\

    \bottomrule
  \end{tabular}
  \caption{\rivet analysis tags, {observables and corresponding
      photon virtuality ranges} used for the tuning.}
  \label{tab:tuning_analysis}
\end{table}

\bibliographystyle{amsunsrt_modp}
\bibliography{references}

\ifx\mcitethebibliography\mciteundefinedmacro
\PackageError{amsunsrt_mod.bst}{mciteplus.sty has not been loaded}
{This bibstyle requires the use of the mciteplus package.}\fi
\begin{mcitethebibliography}{10}

\bibitem{ALEPH:1990iba}
D.~Decamp et~al., The ALEPH collaboration, \emph{{Measurement of the strong
  coupling constant alpha-s from global event shape variables of hadronic Z
  decays}}, Phys. Lett. B \textbf{255} (1991),
  \href{http://inspirehep.net/search?j=Phys%20Lett%20B,255,623}{623--633}%
\relax\mciteBstWouldAddEndPuncttrue
\mciteSetBstMidEndSepPunct{\mcitedefaultmidpunct}
{\mcitedefaultendpunct}{\mcitedefaultseppunct}\relax
\EndOfBibitem
\bibitem{OPAL:1990xiz}
M.~Z. Akrawy et~al., The OPAL collaboration, \emph{{A Measurement of Global
  Event Shape Distributions in the Hadronic Decays of the $Z^0$}}, Z. Phys. C
  \textbf{47} (1990),
  \href{http://inspirehep.net/search?j=Z%20Phys%20C,47,505}{505--522}%
\relax\mciteBstWouldAddEndPuncttrue
\mciteSetBstMidEndSepPunct{\mcitedefaultmidpunct}
{\mcitedefaultendpunct}{\mcitedefaultseppunct}\relax
\EndOfBibitem
\bibitem{L3:1992btq}
O.~Adrian et~al., The L3 collaboration, \emph{{Determination of alpha-s from
  hadronic event shapes measured on the Z0 resonance}}, Phys. Lett. B
  \textbf{284} (1992),
  \href{http://inspirehep.net/search?j=Phys%20Lett%20B,284,471}{471--481}%
\relax\mciteBstWouldAddEndPuncttrue
\mciteSetBstMidEndSepPunct{\mcitedefaultmidpunct}
{\mcitedefaultendpunct}{\mcitedefaultseppunct}\relax
\EndOfBibitem
\bibitem{DELPHI:1999vbd}
P.~Abreu et~al., The DELPHI collaboration, \emph{{Energy dependence of event
  shapes and of alpha(s) at LEP-2}}, Phys. Lett. B \textbf{456} (1999),
  \href{http://inspirehep.net/search?j=Phys%20Lett%20B,456,322}{322--340}%
\relax\mciteBstWouldAddEndPuncttrue
\mciteSetBstMidEndSepPunct{\mcitedefaultmidpunct}
{\mcitedefaultendpunct}{\mcitedefaultseppunct}\relax
\EndOfBibitem
\bibitem{DELPHI:2003yqh}
J.~Abdallah et~al., The DELPHI collaboration, \emph{{A Study of the energy
  evolution of event shape distributions and their means with the DELPHI
  detector at LEP}}, Eur. Phys. J. C \textbf{29} (2003),
  \href{http://inspirehep.net/search?p=hep-ex/0307048}{285--312},
  [\href{http://arXiv.org/pdf/hep-ex/0307048}{{\texttt{hep-ex/0307048}}}]%
\relax\mciteBstWouldAddEndPuncttrue
\mciteSetBstMidEndSepPunct{\mcitedefaultmidpunct}
{\mcitedefaultendpunct}{\mcitedefaultseppunct}\relax
\EndOfBibitem
\bibitem{Marzani:2019hun}
S.~Marzani, G.~Soyez and M.~Spannowsky, \emph{{Looking inside jets: an
  introduction to jet substructure and boosted-object phenomenology}}, vol.
  958, Springer, 2019%
\relax\mciteBstWouldAddEndPuncttrue
\mciteSetBstMidEndSepPunct{\mcitedefaultmidpunct}
{\mcitedefaultendpunct}{\mcitedefaultseppunct}\relax
\EndOfBibitem
\bibitem{H1:1997hbl}
C.~Adloff et~al., The H1 collaboration, \emph{{Measurement of event shape
  variables in deep inelastic e p scattering}}, Phys. Lett. B \textbf{406}
  (1997), \href{http://inspirehep.net/search?p=hep-ex/9706002}{256--270},
  [\href{http://arXiv.org/pdf/hep-ex/9706002}{{\texttt{hep-ex/9706002}}}]%
\relax\mciteBstWouldAddEndPuncttrue
\mciteSetBstMidEndSepPunct{\mcitedefaultmidpunct}
{\mcitedefaultendpunct}{\mcitedefaultseppunct}\relax
\EndOfBibitem
\bibitem{H1:1999wfh}
C.~Adloff et~al., The H1 collaboration, \emph{{Investigation of power
  corrections to event shape variables measured in deep inelastic scattering}},
  Eur. Phys. J. C \textbf{14} (2000),
  \href{http://inspirehep.net/search?p=hep-ex/9912052}{255--269},
  [\href{http://arXiv.org/pdf/hep-ex/9912052}{{\texttt{hep-ex/9912052}}}],
  [Erratum: Eur.Phys.J.C 18, 417--419 (2000)]%
\relax\mciteBstWouldAddEndPuncttrue
\mciteSetBstMidEndSepPunct{\mcitedefaultmidpunct}
{\mcitedefaultendpunct}{\mcitedefaultseppunct}\relax
\EndOfBibitem
\bibitem{H1:2005zsk}
A.~Aktas et~al., The H1 collaboration, \emph{{Measurement of event shape
  variables in deep-inelastic scattering at HERA}}, Eur. Phys. J. C \textbf{46}
  (2006), \href{http://inspirehep.net/search?p=hep-ex/0512014}{343--356},
  [\href{http://arXiv.org/pdf/hep-ex/0512014}{{\texttt{hep-ex/0512014}}}]%
\relax\mciteBstWouldAddEndPuncttrue
\mciteSetBstMidEndSepPunct{\mcitedefaultmidpunct}
{\mcitedefaultendpunct}{\mcitedefaultseppunct}\relax
\EndOfBibitem
\bibitem{ZEUS:1997nib}
J.~Breitweg et~al., The ZEUS collaboration, \emph{{Event shape analysis of deep
  inelastic scattering events with a large rapidity gap at HERA}}, Phys. Lett.
  B \textbf{421} (1998),
  \href{http://inspirehep.net/search?p=hep-ex/9710027}{368--384},
  [\href{http://arXiv.org/pdf/hep-ex/9710027}{{\texttt{hep-ex/9710027}}}]%
\relax\mciteBstWouldAddEndPuncttrue
\mciteSetBstMidEndSepPunct{\mcitedefaultmidpunct}
{\mcitedefaultendpunct}{\mcitedefaultseppunct}\relax
\EndOfBibitem
\bibitem{ZEUS:2002tyf}
S.~Chekanov et~al., The ZEUS collaboration, \emph{{Measurement of event shapes
  in deep inelastic scattering at HERA}}, Eur. Phys. J. C \textbf{27} (2003),
  \href{http://inspirehep.net/search?p=hep-ex/0211040}{531--545},
  [\href{http://arXiv.org/pdf/hep-ex/0211040}{{\texttt{hep-ex/0211040}}}]%
\relax\mciteBstWouldAddEndPuncttrue
\mciteSetBstMidEndSepPunct{\mcitedefaultmidpunct}
{\mcitedefaultendpunct}{\mcitedefaultseppunct}\relax
\EndOfBibitem
\bibitem{ZEUS:2006vwm}
S.~Chekanov et~al., The ZEUS collaboration, \emph{{Event shapes in deep
  inelastic scattering at HERA}}, Nucl. Phys. B \textbf{767} (2007),
  \href{http://inspirehep.net/search?p=hep-ex/0604032}{1--28},
  [\href{http://arXiv.org/pdf/hep-ex/0604032}{{\texttt{hep-ex/0604032}}}]%
\relax\mciteBstWouldAddEndPuncttrue
\mciteSetBstMidEndSepPunct{\mcitedefaultmidpunct}
{\mcitedefaultendpunct}{\mcitedefaultseppunct}\relax
\EndOfBibitem
\bibitem{Schieck:2012mp}
J.~Schieck, S.~Bethke, S.~Kluth, C.~Pahl and Z.~Trocsanyi, The JADE
  collaboration, \emph{{Measurement of the strong coupling $alpha_S$ from the
  three-jet rate in $e^+e^-$ - annihilation using JADE data}}, Eur. Phys. J. C
  \textbf{73} (2013), no.~3,
  \href{http://inspirehep.net/search?p=1205.3714}{2332},
  [\href{http://arXiv.org/pdf/1205.3714}{{\texttt{arXiv:1205.3714}}} [hep-ex]]%
\relax\mciteBstWouldAddEndPuncttrue
\mciteSetBstMidEndSepPunct{\mcitedefaultmidpunct}
{\mcitedefaultendpunct}{\mcitedefaultseppunct}\relax
\EndOfBibitem
\bibitem{ALEPH:2013htx}
G.~Abbiendi et~al., The ALEPH, DELPHI, L3, OPAL, LEP collaboration,
  \emph{{Search for Charged Higgs bosons: Combined Results Using LEP Data}},
  Eur. Phys. J. C \textbf{73} (2013),
  \href{http://inspirehep.net/search?p=1301.6065}{2463},
  [\href{http://arXiv.org/pdf/1301.6065}{{\texttt{arXiv:1301.6065}}} [hep-ex]]%
\relax\mciteBstWouldAddEndPuncttrue
\mciteSetBstMidEndSepPunct{\mcitedefaultmidpunct}
{\mcitedefaultendpunct}{\mcitedefaultseppunct}\relax
\EndOfBibitem
\bibitem{DELPHI:2014nkw}
J.~Abdallah et~al., The DELPHI collaboration, \emph{{Measurement of the
  electron structure function F$\frac{e}{2}$ at LEP energies}}, Phys. Lett. B
  \textbf{737} (2014),
  \href{http://inspirehep.net/search?j=Phys%20Lett%20B,737,39}{39--47}%
\relax\mciteBstWouldAddEndPuncttrue
\mciteSetBstMidEndSepPunct{\mcitedefaultmidpunct}
{\mcitedefaultendpunct}{\mcitedefaultseppunct}\relax
\EndOfBibitem
\bibitem{Fischer:2015pqa}
N.~Fischer, S.~Gieseke, S.~Kluth, S.~Pl\"atzer and P.~Skands, The OPAL
  collaboration, \emph{{Measurement of observables sensitive to coherence
  effects in hadronic Z decays with the OPAL detector at LEP}}, Eur. Phys. J. C
  \textbf{75} (2015), no.~12,
  \href{http://inspirehep.net/search?p=1505.01636}{571},
  [\href{http://arXiv.org/pdf/1505.01636}{{\texttt{arXiv:1505.01636}}}
  [hep-ex]]%
\relax\mciteBstWouldAddEndPuncttrue
\mciteSetBstMidEndSepPunct{\mcitedefaultmidpunct}
{\mcitedefaultendpunct}{\mcitedefaultseppunct}\relax
\EndOfBibitem
\bibitem{Badea:2019vey}
A.~Badea, A.~Baty, P.~Chang, G.~M. Innocenti, M.~Maggi, C.~Mcginn, M.~Peters,
  T.-A. Sheng, J.~Thaler and Y.-J. Lee, \emph{{Measurements of two-particle
  correlations in $e^+e^-$ collisions at 91 GeV with ALEPH archived data}},
  Phys. Rev. Lett. \textbf{123} (2019), no.~21,
  \href{http://inspirehep.net/search?p=1906.00489}{212002},
  [\href{http://arXiv.org/pdf/1906.00489}{{\texttt{arXiv:1906.00489}}}
  [hep-ex]]%
\relax\mciteBstWouldAddEndPuncttrue
\mciteSetBstMidEndSepPunct{\mcitedefaultmidpunct}
{\mcitedefaultendpunct}{\mcitedefaultseppunct}\relax
\EndOfBibitem
\bibitem{Chen:2021uws}
Y.~Chen et~al., \emph{{Jet energy spectrum and substructure in e$^{+}$e$^{-}$
  collisions at 91.2 GeV with ALEPH Archived Data}}, JHEP \textbf{06} (2022),
  \href{http://inspirehep.net/search?p=2111.09914}{008},
  [\href{http://arXiv.org/pdf/2111.09914}{{\texttt{arXiv:2111.09914}}}
  [hep-ex]]%
\relax\mciteBstWouldAddEndPuncttrue
\mciteSetBstMidEndSepPunct{\mcitedefaultmidpunct}
{\mcitedefaultendpunct}{\mcitedefaultseppunct}\relax
\EndOfBibitem
\bibitem{Accardi:2012qut}
A.~Accardi et~al., \emph{{Electron Ion Collider: The Next QCD Frontier}:
  {Understanding the glue that binds us all}}, Eur. Phys. J. A \textbf{52}
  (2016), no.~9, \href{http://inspirehep.net/search?p=1212.1701}{268},
  [\href{http://arXiv.org/pdf/1212.1701}{{\texttt{arXiv:1212.1701}}}
  [nucl-ex]]%
\relax\mciteBstWouldAddEndPuncttrue
\mciteSetBstMidEndSepPunct{\mcitedefaultmidpunct}
{\mcitedefaultendpunct}{\mcitedefaultseppunct}\relax
\EndOfBibitem
\bibitem{AbdulKhalek:2021gbh}
R.~Abdul~Khalek et~al., \emph{{Science Requirements and Detector Concepts for
  the Electron-Ion Collider}: {EIC Yellow Report}}, Nucl. Phys. A \textbf{1026}
  (2022), \href{http://inspirehep.net/search?p=2103.05419}{122447},
  [\href{http://arXiv.org/pdf/2103.05419}{{\texttt{arXiv:2103.05419}}}
  [physics.ins-det]]%
\relax\mciteBstWouldAddEndPuncttrue
\mciteSetBstMidEndSepPunct{\mcitedefaultmidpunct}
{\mcitedefaultendpunct}{\mcitedefaultseppunct}\relax
\EndOfBibitem
\bibitem{LHeCStudyGroup:2012zhm}
J.~L. Abelleira~Fernandez et~al., The LHeC Study Group collaboration, \emph{{A
  Large Hadron Electron Collider at CERN: Report on the Physics and Design
  Concepts for Machine and Detector}}, J. Phys. G \textbf{39} (2012),
  \href{http://inspirehep.net/search?p=1206.2913}{075001},
  [\href{http://arXiv.org/pdf/1206.2913}{{\texttt{arXiv:1206.2913}}}
  [physics.acc-ph]]%
\relax\mciteBstWouldAddEndPuncttrue
\mciteSetBstMidEndSepPunct{\mcitedefaultmidpunct}
{\mcitedefaultendpunct}{\mcitedefaultseppunct}\relax
\EndOfBibitem
\bibitem{LHeC:2020van}
P.~Agostini et~al., The LHeC, FCC-he Study Group collaboration, \emph{{The
  Large Hadron\textendash{}Electron Collider at the HL-LHC}}, J. Phys. G
  \textbf{48} (2021), no.~11,
  \href{http://inspirehep.net/search?p=2007.14491}{110501},
  [\href{http://arXiv.org/pdf/2007.14491}{{\texttt{arXiv:2007.14491}}}
  [hep-ex]]%
\relax\mciteBstWouldAddEndPuncttrue
\mciteSetBstMidEndSepPunct{\mcitedefaultmidpunct}
{\mcitedefaultendpunct}{\mcitedefaultseppunct}\relax
\EndOfBibitem
\bibitem{Kang:2013nha}
D.~Kang, C.~Lee and I.~W. Stewart, \emph{{Using 1-Jettiness to Measure 2 Jets
  in DIS 3 Ways}}, Phys. Rev. D \textbf{88} (2013),
  \href{http://inspirehep.net/search?p=1303.6952}{054004},
  [\href{http://arXiv.org/pdf/1303.6952}{{\texttt{arXiv:1303.6952}}} [hep-ph]]%
\relax\mciteBstWouldAddEndPuncttrue
\mciteSetBstMidEndSepPunct{\mcitedefaultmidpunct}
{\mcitedefaultendpunct}{\mcitedefaultseppunct}\relax
\EndOfBibitem
\bibitem{Antonelli:1999kx}
V.~Antonelli, M.~Dasgupta and G.~P. Salam, \emph{{Resummation of thrust
  distributions in DIS}}, JHEP \textbf{02} (2000),
  \href{http://inspirehep.net/search?p=hep-ph/9912488}{001},
  [\href{http://arXiv.org/pdf/hep-ph/9912488}{{\texttt{hep-ph/9912488}}}]%
\relax\mciteBstWouldAddEndPuncttrue
\mciteSetBstMidEndSepPunct{\mcitedefaultmidpunct}
{\mcitedefaultendpunct}{\mcitedefaultseppunct}\relax
\EndOfBibitem
\bibitem{Chahal:2022rid}
G.~S. Chahal and F.~Krauss, \emph{{Cluster Hadronisation in Sherpa}}, SciPost
  Phys. \textbf{13} (2022), no.~2,
  \href{http://inspirehep.net/search?p=2203.11385}{019},
  [\href{http://arXiv.org/pdf/2203.11385}{{\texttt{arXiv:2203.11385}}}
  [hep-ph]]%
\relax\mciteBstWouldAddEndPuncttrue
\mciteSetBstMidEndSepPunct{\mcitedefaultmidpunct}
{\mcitedefaultendpunct}{\mcitedefaultseppunct}\relax
\EndOfBibitem
\bibitem{Knobbe:2023njd}
\href{http://inspirehep.net/search?p=2306.03682}{M.~Knobbe, F.~Krauss,
  D.~Reichelt and S.~Schumann}, \emph{{Measuring Hadronic Higgs Boson Branching
  Ratios at Future Lepton Colliders}},
  \href{http://arXiv.org/pdf/2306.03682}{{\texttt{arXiv:2306.03682}}} [hep-ph]%
\relax\mciteBstWouldAddEndPuncttrue
\mciteSetBstMidEndSepPunct{\mcitedefaultmidpunct}
{\mcitedefaultendpunct}{\mcitedefaultseppunct}\relax
\EndOfBibitem
\bibitem{Banfi:2004yd}
A.~Banfi, G.~P. Salam and G.~Zanderighi, \emph{{Principles of general
  final-state resummation and automated implementation}}, JHEP \textbf{03}
  (2005), \href{http://inspirehep.net/search?p=hep-ph/0407286}{073},
  [\href{http://arXiv.org/pdf/hep-ph/0407286}{{\texttt{hep-ph/0407286}}}]%
\relax\mciteBstWouldAddEndPuncttrue
\mciteSetBstMidEndSepPunct{\mcitedefaultmidpunct}
{\mcitedefaultendpunct}{\mcitedefaultseppunct}\relax
\EndOfBibitem
\bibitem{Gerwick:2014gya}
E.~Gerwick, S.~H{\"o}che, S.~Marzani and S.~Schumann, \emph{{Soft evolution of
  multi-jet final states}}, JHEP \textbf{02} (2015),
  \href{http://inspirehep.net/search?p=1411.7325}{106},
  [\href{http://arXiv.org/pdf/1411.7325}{{\texttt{arXiv:1411.7325}}} [hep-ph]]%
\relax\mciteBstWouldAddEndPuncttrue
\mciteSetBstMidEndSepPunct{\mcitedefaultmidpunct}
{\mcitedefaultendpunct}{\mcitedefaultseppunct}\relax
\EndOfBibitem
\bibitem{Hoche:2018gti}
S.~H\"oche, S.~Kuttimalai and Y.~Li, \emph{{Hadronic Final States in DIS at
  NNLO QCD with Parton Showers}}, Phys. Rev. D \textbf{98} (2018), no.~11,
  \href{http://inspirehep.net/search?p=1809.04192}{114013},
  [\href{http://arXiv.org/pdf/1809.04192}{{\texttt{arXiv:1809.04192}}}
  [hep-ph]]%
\relax\mciteBstWouldAddEndPuncttrue
\mciteSetBstMidEndSepPunct{\mcitedefaultmidpunct}
{\mcitedefaultendpunct}{\mcitedefaultseppunct}\relax
\EndOfBibitem
\bibitem{Reichelt:2021svh}
D.~Reichelt, S.~Caletti, O.~Fedkevych, S.~Marzani, S.~Schumann and G.~Soyez,
  \emph{{Phenomenology of jet angularities at the LHC}}, JHEP \textbf{03}
  (2022), \href{http://inspirehep.net/search?p=2112.09545}{131},
  [\href{http://arXiv.org/pdf/2112.09545}{{\texttt{arXiv:2112.09545}}}
  [hep-ph]]%
\relax\mciteBstWouldAddEndPuncttrue
\mciteSetBstMidEndSepPunct{\mcitedefaultmidpunct}
{\mcitedefaultendpunct}{\mcitedefaultseppunct}\relax
\EndOfBibitem
\bibitem{Hessler:2021usr}
J.~Hessler, D.~Britzger and S.~Lee, The H1 collaboration, \emph{{Measurement of
  1-jettiness in deep-inelastic scattering at HERA}}, PoS \textbf{EPS-HEP2021}
  (2022), \href{http://inspirehep.net/search?p=2111.11364}{367},
  [\href{http://arXiv.org/pdf/2111.11364}{{\texttt{arXiv:2111.11364}}}
  [hep-ex]]%
\relax\mciteBstWouldAddEndPuncttrue
\mciteSetBstMidEndSepPunct{\mcitedefaultmidpunct}
{\mcitedefaultendpunct}{\mcitedefaultseppunct}\relax
\EndOfBibitem
\bibitem{Hessler:2021ubp}
J.~Hessler, \emph{{Measurement of the 1-jettiness Event Shape Observable in
  Deep-inelastic Electron-Proton Scattering}}, Master's thesis, Munich, Tech.
  U., 2021%
\relax\mciteBstWouldAddEndPuncttrue
\mciteSetBstMidEndSepPunct{\mcitedefaultmidpunct}
{\mcitedefaultendpunct}{\mcitedefaultseppunct}\relax
\EndOfBibitem
\bibitem{H1:2023fzk}
\href{http://inspirehep.net/search?p=2303.13620}{V.~Andreev et~al.}, The H1
  collaboration, \emph{{Unbinned Deep Learning Jet Substructure Measurement in
  High $Q^2$ ep collisions at HERA}},
  \href{http://arXiv.org/pdf/2303.13620}{{\texttt{arXiv:2303.13620}}} [hep-ex]%
\relax\mciteBstWouldAddEndPuncttrue
\mciteSetBstMidEndSepPunct{\mcitedefaultmidpunct}
{\mcitedefaultendpunct}{\mcitedefaultseppunct}\relax
\EndOfBibitem
\bibitem{Carli:2010cg}
T.~Carli, T.~Gehrmann and S.~H{\"o}che, \emph{{Hadronic final states in
  deep-inelastic scattering with Sherpa}}, Eur. Phys. J. C \textbf{67} (2010),
  \href{http://inspirehep.net/search?p=0912.3715}{73--97},
  [\href{http://arXiv.org/pdf/0912.3715}{{\texttt{arXiv:0912.3715}}} [hep-ph]]%
\relax\mciteBstWouldAddEndPuncttrue
\mciteSetBstMidEndSepPunct{\mcitedefaultmidpunct}
{\mcitedefaultendpunct}{\mcitedefaultseppunct}\relax
\EndOfBibitem
\bibitem{Larkoski:2014wba}
A.~J. Larkoski, S.~Marzani, G.~Soyez and J.~Thaler, \emph{{Soft Drop}}, JHEP
  \textbf{05} (2014), \href{http://inspirehep.net/search?p=1402.2657}{146},
  [\href{http://arXiv.org/pdf/1402.2657}{{\texttt{arXiv:1402.2657}}} [hep-ph]]%
\relax\mciteBstWouldAddEndPuncttrue
\mciteSetBstMidEndSepPunct{\mcitedefaultmidpunct}
{\mcitedefaultendpunct}{\mcitedefaultseppunct}\relax
\EndOfBibitem
\bibitem{Butterworth:2008iy}
J.~M. Butterworth, A.~R. Davison, M.~Rubin and G.~P. Salam, \emph{{Jet
  substructure as a new Higgs search channel at the LHC}}, Phys. Rev. Lett.
  \textbf{100} (2008), \href{http://inspirebeta.net/record/779602}{242001},
  [\href{http://arXiv.org/pdf/0802.2470}{{\texttt{arXiv:0802.2470}}} [hep-ph]]%
\relax\mciteBstWouldAddEndPuncttrue
\mciteSetBstMidEndSepPunct{\mcitedefaultmidpunct}
{\mcitedefaultendpunct}{\mcitedefaultseppunct}\relax
\EndOfBibitem
\bibitem{Dasgupta:2013ihk}
M.~Dasgupta, A.~Fregoso, S.~Marzani and G.~P. Salam, \emph{{Towards an
  understanding of jet substructure}}, JHEP \textbf{1309} (2013),
  \href{http://inspirehep.net/search?p=1307.0007}{029},
  [\href{http://arXiv.org/pdf/1307.0007}{{\texttt{arXiv:1307.0007}}} [hep-ph]]%
\relax\mciteBstWouldAddEndPuncttrue
\mciteSetBstMidEndSepPunct{\mcitedefaultmidpunct}
{\mcitedefaultendpunct}{\mcitedefaultseppunct}\relax
\EndOfBibitem
\bibitem{Baron:2018nfz}
J.~Baron, S.~Marzani and V.~Theeuwes, \emph{{Soft-Drop Thrust}}, JHEP
  \textbf{08} (2018), \href{http://inspirehep.net/search?p=1803.04719}{105},
  [\href{http://arXiv.org/pdf/1803.04719}{{\texttt{arXiv:1803.04719}}}
  [hep-ph]], [Erratum: JHEP 05, 056 (2019)]%
\relax\mciteBstWouldAddEndPuncttrue
\mciteSetBstMidEndSepPunct{\mcitedefaultmidpunct}
{\mcitedefaultendpunct}{\mcitedefaultseppunct}\relax
\EndOfBibitem
\bibitem{Marzani:2019evv}
S.~Marzani, D.~Reichelt, S.~Schumann, G.~Soyez and V.~Theeuwes, \emph{{Fitting
  the Strong Coupling Constant with Soft-Drop Thrust}}, JHEP \textbf{11}
  (2019), \href{http://inspirehep.net/search?p=1906.10504}{179},
  [\href{http://arXiv.org/pdf/1906.10504}{{\texttt{arXiv:1906.10504}}}
  [hep-ph]]%
\relax\mciteBstWouldAddEndPuncttrue
\mciteSetBstMidEndSepPunct{\mcitedefaultmidpunct}
{\mcitedefaultendpunct}{\mcitedefaultseppunct}\relax
\EndOfBibitem
\bibitem{Baron:2020xoi}
J.~Baron, D.~Reichelt, S.~Schumann, N.~Schwanemann and V.~Theeuwes,
  \emph{{Soft-drop grooming for hadronic event shapes}}, JHEP \textbf{07}
  (2021), \href{http://inspirehep.net/search?p=2012.09574}{142},
  [\href{http://arXiv.org/pdf/2012.09574}{{\texttt{arXiv:2012.09574}}}
  [hep-ph]]%
\relax\mciteBstWouldAddEndPuncttrue
\mciteSetBstMidEndSepPunct{\mcitedefaultmidpunct}
{\mcitedefaultendpunct}{\mcitedefaultseppunct}\relax
\EndOfBibitem
\bibitem{Makris:2021drz}
Y.~Makris, \emph{{Revisiting the role of grooming in DIS}}, Phys. Rev. D
  \textbf{103} (2021), no.~5,
  \href{http://inspirehep.net/search?p=2101.02708}{054005},
  [\href{http://arXiv.org/pdf/2101.02708}{{\texttt{arXiv:2101.02708}}}
  [hep-ph]]%
\relax\mciteBstWouldAddEndPuncttrue
\mciteSetBstMidEndSepPunct{\mcitedefaultmidpunct}
{\mcitedefaultendpunct}{\mcitedefaultseppunct}\relax
\EndOfBibitem
\bibitem{Arratia:2020ssx}
M.~Arratia, Y.~Makris, D.~Neill, F.~Ringer and N.~Sato, \emph{{Asymmetric jet
  clustering in deep-inelastic scattering}}, Phys. Rev. D \textbf{104} (2021),
  no.~3, \href{http://inspirehep.net/search?p=2006.10751}{034005},
  [\href{http://arXiv.org/pdf/2006.10751}{{\texttt{arXiv:2006.10751}}}
  [hep-ph]]%
\relax\mciteBstWouldAddEndPuncttrue
\mciteSetBstMidEndSepPunct{\mcitedefaultmidpunct}
{\mcitedefaultendpunct}{\mcitedefaultseppunct}\relax
\EndOfBibitem
\bibitem{Sherpa3.0.beta}
\emph{\normalfont{The \sherpa-3.0.beta code can be obtained from:}
  \url{https://sherpa-team.gitlab.io/changelog.html}}%
\relax\mciteBstWouldAddEndPuncttrue
\mciteSetBstMidEndSepPunct{\mcitedefaultmidpunct}
{\mcitedefaultendpunct}{\mcitedefaultseppunct}\relax
\EndOfBibitem
\bibitem{Sherpa:2019gpd}
E.~Bothmann et~al., The Sherpa collaboration, \emph{{Event Generation with
  Sherpa 2.2}}, SciPost Phys. \textbf{7} (2019), no.~3,
  \href{http://inspirehep.net/search?p=1905.09127}{034},
  [\href{http://arXiv.org/pdf/1905.09127}{{\texttt{arXiv:1905.09127}}}
  [hep-ph]]%
\relax\mciteBstWouldAddEndPuncttrue
\mciteSetBstMidEndSepPunct{\mcitedefaultmidpunct}
{\mcitedefaultendpunct}{\mcitedefaultseppunct}\relax
\EndOfBibitem
\bibitem{Schonherr:2017qcj}
M.~Sch\"onherr, \emph{{An automated subtraction of NLO EW infrared
  divergences}}, Eur. Phys. J. C \textbf{78} (2018), no.~2,
  \href{http://inspirehep.net/search?p=1712.07975}{119},
  [\href{http://arXiv.org/pdf/1712.07975}{{\texttt{arXiv:1712.07975}}}
  [hep-ph]]%
\relax\mciteBstWouldAddEndPuncttrue
\mciteSetBstMidEndSepPunct{\mcitedefaultmidpunct}
{\mcitedefaultendpunct}{\mcitedefaultseppunct}\relax
\EndOfBibitem
\bibitem{Schonherr:2018jva}
M.~Sch\"onherr, \emph{{Next-to-leading order electroweak corrections to
  off-shell WWW production at the LHC}}, JHEP \textbf{07} (2018),
  \href{http://inspirehep.net/search?p=1806.00307}{076},
  [\href{http://arXiv.org/pdf/1806.00307}{{\texttt{arXiv:1806.00307}}}
  [hep-ph]]%
\relax\mciteBstWouldAddEndPuncttrue
\mciteSetBstMidEndSepPunct{\mcitedefaultmidpunct}
{\mcitedefaultendpunct}{\mcitedefaultseppunct}\relax
\EndOfBibitem
\bibitem{Brauer:2020kfv}
S.~Br\"auer, A.~Denner, M.~Pellen, M.~Sch\"onherr and S.~Schumann,
  \emph{{Fixed-order and merged parton-shower predictions for WW and WWj
  production at the LHC including NLO QCD and EW corrections}}, JHEP
  \textbf{10} (2020), \href{http://inspirehep.net/search?p=2005.12128}{159},
  [\href{http://arXiv.org/pdf/2005.12128}{{\texttt{arXiv:2005.12128}}}
  [hep-ph]]%
\relax\mciteBstWouldAddEndPuncttrue
\mciteSetBstMidEndSepPunct{\mcitedefaultmidpunct}
{\mcitedefaultendpunct}{\mcitedefaultseppunct}\relax
\EndOfBibitem
\bibitem{Bothmann:2020sxm}
E.~Bothmann and D.~Napoletano, \emph{{Automated evaluation of electroweak
  Sudakov logarithms in Sherpa}}, Eur. Phys. J. C \textbf{80} (2020), no.~11,
  \href{http://inspirehep.net/search?p=2006.14635}{1024},
  [\href{http://arXiv.org/pdf/2006.14635}{{\texttt{arXiv:2006.14635}}}
  [hep-ph]]%
\relax\mciteBstWouldAddEndPuncttrue
\mciteSetBstMidEndSepPunct{\mcitedefaultmidpunct}
{\mcitedefaultendpunct}{\mcitedefaultseppunct}\relax
\EndOfBibitem
\bibitem{Bothmann:2021led}
E.~Bothmann, D.~Napoletano, M.~Sch\"onherr, S.~Schumann and S.~L. Villani,
  \emph{{Higher-order EW corrections in ZZ and ZZj production at the LHC}},
  JHEP \textbf{06} (2022),
  \href{http://inspirehep.net/search?p=2111.13453}{064},
  [\href{http://arXiv.org/pdf/2111.13453}{{\texttt{arXiv:2111.13453}}}
  [hep-ph]]%
\relax\mciteBstWouldAddEndPuncttrue
\mciteSetBstMidEndSepPunct{\mcitedefaultmidpunct}
{\mcitedefaultendpunct}{\mcitedefaultseppunct}\relax
\EndOfBibitem
\bibitem{yaml}
\href{https://yaml.org/}{} \url{https://yaml.org/}%
\relax\mciteBstWouldAddEndPuncttrue
\mciteSetBstMidEndSepPunct{\mcitedefaultmidpunct}
{\mcitedefaultendpunct}{\mcitedefaultseppunct}\relax
\EndOfBibitem
\bibitem{Bierlich:2019rhm}
C.~Bierlich et~al., \emph{{Robust Independent Validation of Experiment and
  Theory: Rivet version 3}}, SciPost Phys. \textbf{8} (2020),
  \href{http://inspirehep.net/search?p=1912.05451}{026},
  [\href{http://arXiv.org/pdf/1912.05451}{{\texttt{arXiv:1912.05451}}}
  [hep-ph]]%
\relax\mciteBstWouldAddEndPuncttrue
\mciteSetBstMidEndSepPunct{\mcitedefaultmidpunct}
{\mcitedefaultendpunct}{\mcitedefaultseppunct}\relax
\EndOfBibitem
\bibitem{Cacciari:2011ma}
M.~Cacciari, G.~P. Salam and G.~Soyez, \emph{{FastJet User Manual}}, Eur. Phys.
  J. \textbf{C72} (2012),
  \href{http://inspirehep.net/search?p=1111.6097}{1896},
  [\href{http://arXiv.org/pdf/1111.6097}{{\texttt{arXiv:1111.6097}}} [hep-ph]]%
\relax\mciteBstWouldAddEndPuncttrue
\mciteSetBstMidEndSepPunct{\mcitedefaultmidpunct}
{\mcitedefaultendpunct}{\mcitedefaultseppunct}\relax
\EndOfBibitem
\bibitem{Schumann:2007mg}
S.~Schumann and F.~Krauss, \emph{{A Parton shower algorithm based on
  Catani-Seymour dipole factorisation}}, JHEP \textbf{03} (2008),
  \href{http://inspirehep.net/search?p=0709.1027}{038},
  [\href{http://arXiv.org/pdf/0709.1027}{{\texttt{arXiv:0709.1027}}} [hep-ph]]%
\relax\mciteBstWouldAddEndPuncttrue
\mciteSetBstMidEndSepPunct{\mcitedefaultmidpunct}
{\mcitedefaultendpunct}{\mcitedefaultseppunct}\relax
\EndOfBibitem
\bibitem{Hoeche:2012yf}
S.~H{\"o}che, F.~Krauss, M.~Sch{\"o}nherr and F.~Siegert, \emph{{QCD matrix
  elements + parton showers: The NLO case}}, JHEP \textbf{04} (2013),
  \href{http://inspirehep.net/search?p=1207.5030}{027},
  [\href{http://arXiv.org/pdf/1207.5030}{{\texttt{arXiv:1207.5030}}} [hep-ph]]%
\relax\mciteBstWouldAddEndPuncttrue
\mciteSetBstMidEndSepPunct{\mcitedefaultmidpunct}
{\mcitedefaultendpunct}{\mcitedefaultseppunct}\relax
\EndOfBibitem
\bibitem{Hoeche:2009rj}
S.~H{\"o}che, F.~Krauss, S.~Schumann and F.~Siegert, \emph{{QCD matrix elements
  and truncated showers}}, JHEP \textbf{05} (2009),
  \href{http://inspirehep.net/search?p=0903.1219}{053},
  [\href{http://arXiv.org/pdf/0903.1219}{{\texttt{arXiv:0903.1219}}} [hep-ph]]%
\relax\mciteBstWouldAddEndPuncttrue
\mciteSetBstMidEndSepPunct{\mcitedefaultmidpunct}
{\mcitedefaultendpunct}{\mcitedefaultseppunct}\relax
\EndOfBibitem
\bibitem{Buccioni:2019sur}
F.~Buccioni, J.-N. Lang, J.~M. Lindert, P.~Maierh\"ofer, S.~Pozzorini, H.~Zhang
  and M.~F. Zoller, \emph{{OpenLoops 2}}, Eur. Phys. J. C \textbf{79} (2019),
  no.~10, \href{http://inspirehep.net/search?p=1907.13071}{866},
  [\href{http://arXiv.org/pdf/1907.13071}{{\texttt{arXiv:1907.13071}}}
  [hep-ph]]%
\relax\mciteBstWouldAddEndPuncttrue
\mciteSetBstMidEndSepPunct{\mcitedefaultmidpunct}
{\mcitedefaultendpunct}{\mcitedefaultseppunct}\relax
\EndOfBibitem
\bibitem{Denner:2016kdg}
A.~Denner, S.~Dittmaier and L.~Hofer, \emph{{Collier: a fortran-based Complex
  One-Loop LIbrary in Extended Regularizations}}, Comput. Phys. Commun.
  \textbf{212} (2017),
  \href{http://inspirehep.net/search?p=1604.06792}{220--238},
  [\href{http://arXiv.org/pdf/1604.06792}{{\texttt{arXiv:1604.06792}}}
  [hep-ph]]%
\relax\mciteBstWouldAddEndPuncttrue
\mciteSetBstMidEndSepPunct{\mcitedefaultmidpunct}
{\mcitedefaultendpunct}{\mcitedefaultseppunct}\relax
\EndOfBibitem
\bibitem{Gleisberg:2008fv}
T.~Gleisberg and S.~H{\"o}che, \emph{{Comix, a new matrix element generator}},
  JHEP \textbf{12} (2008),
  \href{http://inspirehep.net/search?p=0808.3674}{039},
  [\href{http://arXiv.org/pdf/0808.3674}{{\texttt{arXiv:0808.3674}}} [hep-ph]]%
\relax\mciteBstWouldAddEndPuncttrue
\mciteSetBstMidEndSepPunct{\mcitedefaultmidpunct}
{\mcitedefaultendpunct}{\mcitedefaultseppunct}\relax
\EndOfBibitem
\bibitem{Buckley:2014ana}
A.~Buckley, J.~Ferrando, S.~Lloyd, K.~Nordstr\"om, B.~Page, M.~R\"ufenacht,
  M.~Sch\"onherr and G.~Watt, \emph{{LHAPDF6: parton density access in the LHC
  precision era}}, Eur. Phys. J. C \textbf{75} (2015),
  \href{http://inspirehep.net/search?p=1412.7420}{132},
  [\href{http://arXiv.org/pdf/1412.7420}{{\texttt{arXiv:1412.7420}}} [hep-ph]]%
\relax\mciteBstWouldAddEndPuncttrue
\mciteSetBstMidEndSepPunct{\mcitedefaultmidpunct}
{\mcitedefaultendpunct}{\mcitedefaultseppunct}\relax
\EndOfBibitem
\bibitem{PDF4LHCWorkingGroup:2022cjn}
R.~D. Ball et~al., The PDF4LHC Working Group collaboration, \emph{{The
  PDF4LHC21 combination of global PDF fits for the LHC Run III}}, J. Phys. G
  \textbf{49} (2022), no.~8,
  \href{http://inspirehep.net/search?p=2203.05506}{080501},
  [\href{http://arXiv.org/pdf/2203.05506}{{\texttt{arXiv:2203.05506}}}
  [hep-ph]]%
\relax\mciteBstWouldAddEndPuncttrue
\mciteSetBstMidEndSepPunct{\mcitedefaultmidpunct}
{\mcitedefaultendpunct}{\mcitedefaultseppunct}\relax
\EndOfBibitem
\bibitem{Bothmann:2016nao}
E.~Bothmann, M.~Sch\"onherr and S.~Schumann, \emph{{Reweighting QCD
  matrix-element and parton-shower calculations}}, Eur. Phys. J. C \textbf{76}
  (2016), no.~11, \href{http://inspirehep.net/search?p=1606.08753}{590},
  [\href{http://arXiv.org/pdf/1606.08753}{{\texttt{arXiv:1606.08753}}}
  [hep-ph]]%
\relax\mciteBstWouldAddEndPuncttrue
\mciteSetBstMidEndSepPunct{\mcitedefaultmidpunct}
{\mcitedefaultendpunct}{\mcitedefaultseppunct}\relax
\EndOfBibitem
\bibitem{Winter:2003tt}
J.-C. Winter, F.~Krauss and G.~Soff, \emph{{A Modified cluster hadronization
  model}}, Eur. Phys. J. C \textbf{36} (2004),
  \href{http://inspirehep.net/search?p=hep-ph/0311085}{381--395},
  [\href{http://arXiv.org/pdf/hep-ph/0311085}{{\texttt{hep-ph/0311085}}}]%
\relax\mciteBstWouldAddEndPuncttrue
\mciteSetBstMidEndSepPunct{\mcitedefaultmidpunct}
{\mcitedefaultendpunct}{\mcitedefaultseppunct}\relax
\EndOfBibitem
\bibitem{Gleisberg:2008ta}
T.~Gleisberg, S.~H{\"o}che, F.~Krauss, M.~Sch{\"o}nherr, S.~Schumann,
  F.~Siegert and J.~Winter, \emph{{Event generation with SHERPA 1.1}}, JHEP
  \textbf{02} (2009), \href{http://inspirehep.net/search?p=0811.4622}{007},
  [\href{http://arXiv.org/pdf/0811.4622}{{\texttt{arXiv:0811.4622}}} [hep-ph]]%
\relax\mciteBstWouldAddEndPuncttrue
\mciteSetBstMidEndSepPunct{\mcitedefaultmidpunct}
{\mcitedefaultendpunct}{\mcitedefaultseppunct}\relax
\EndOfBibitem
\bibitem{Marchesini:1983bm}
G.~Marchesini and B.~R. Webber, \emph{{Simulation of QCD Jets Including Soft
  Gluon Interference}}, Nucl. Phys. B \textbf{238} (1984),
  \href{http://inspirehep.net/search?j=Nucl%20Phys%20B,238,1}{1--29}%
\relax\mciteBstWouldAddEndPuncttrue
\mciteSetBstMidEndSepPunct{\mcitedefaultmidpunct}
{\mcitedefaultendpunct}{\mcitedefaultseppunct}\relax
\EndOfBibitem
\bibitem{Webber:1983if}
B.~R. Webber, \emph{{A QCD Model for Jet Fragmentation Including Soft Gluon
  Interference}}, Nucl. Phys. B \textbf{238} (1984),
  \href{http://inspirehep.net/search?j=Nucl%20Phys%20B,238,492}{492--528}%
\relax\mciteBstWouldAddEndPuncttrue
\mciteSetBstMidEndSepPunct{\mcitedefaultmidpunct}
{\mcitedefaultendpunct}{\mcitedefaultseppunct}\relax
\EndOfBibitem
\bibitem{Amati:1979fg}
D.~Amati and G.~Veneziano, \emph{{Preconfinement as a Property of Perturbative
  QCD}}, Phys. Lett. B \textbf{83} (1979),
  \href{http://inspirehep.net/search?j=Phys%20Lett%20B,83,87}{87--92}%
\relax\mciteBstWouldAddEndPuncttrue
\mciteSetBstMidEndSepPunct{\mcitedefaultmidpunct}
{\mcitedefaultendpunct}{\mcitedefaultseppunct}\relax
\EndOfBibitem
\bibitem{Andersson:1983ia}
B.~Andersson, G.~Gustafson, G.~Ingelman and T.~Sjostrand, \emph{{Parton
  Fragmentation and String Dynamics}}, Phys. Rept. \textbf{97} (1983),
  \href{http://inspirehep.net/search?j=Phys%20Rept,97,31}{31--145}%
\relax\mciteBstWouldAddEndPuncttrue
\mciteSetBstMidEndSepPunct{\mcitedefaultmidpunct}
{\mcitedefaultendpunct}{\mcitedefaultseppunct}\relax
\EndOfBibitem
\bibitem{Krishnamoorthy:2021nwv}
M.~Krishnamoorthy, H.~Schulz, X.~Ju, W.~Wang, S.~Leyffer, Z.~Marshall,
  S.~Mrenna, J.~M\"uller and J.~B. Kowalkowski, \emph{{Apprentice for Event
  Generator Tuning}}, EPJ Web Conf. \textbf{251} (2021),
  \href{http://inspirehep.net/search?p=2103.05748}{03060},
  [\href{http://arXiv.org/pdf/2103.05748}{{\texttt{arXiv:2103.05748}}}
  [hep-ex]]%
\relax\mciteBstWouldAddEndPuncttrue
\mciteSetBstMidEndSepPunct{\mcitedefaultmidpunct}
{\mcitedefaultendpunct}{\mcitedefaultseppunct}\relax
\EndOfBibitem
\bibitem{H1:1999dbb}
C.~Adloff et~al., The H1 collaboration, \emph{{Measurements of transverse
  energy flow in deep inelastic scattering at HERA}}, Eur. Phys. J. C
  \textbf{12} (2000),
  \href{http://inspirehep.net/search?p=hep-ex/9907027}{595--607},
  [\href{http://arXiv.org/pdf/hep-ex/9907027}{{\texttt{hep-ex/9907027}}}]%
\relax\mciteBstWouldAddEndPuncttrue
\mciteSetBstMidEndSepPunct{\mcitedefaultmidpunct}
{\mcitedefaultendpunct}{\mcitedefaultseppunct}\relax
\EndOfBibitem
\bibitem{H1:1994vjx}
I.~Abt et~al., The H1 collaboration, \emph{{Energy flow and charged particle
  spectrum in deep inelastic scattering at HERA}}, Z. Phys. C \textbf{63}
  (1994), \href{http://inspirehep.net/search?j=Z%20Phys%20C,63,377}{377--390}%
\relax\mciteBstWouldAddEndPuncttrue
\mciteSetBstMidEndSepPunct{\mcitedefaultmidpunct}
{\mcitedefaultendpunct}{\mcitedefaultseppunct}\relax
\EndOfBibitem
\bibitem{ZEUS:1995red}
M.~Derrick et~al., The ZEUS collaboration, \emph{{Measurement of multiplicity
  and momentum spectra in the current fragmentation region of the Breit frame
  at HERA}}, Z. Phys. C \textbf{67} (1995),
  \href{http://inspirehep.net/search?p=hep-ex/9501012}{93--108},
  [\href{http://arXiv.org/pdf/hep-ex/9501012}{{\texttt{hep-ex/9501012}}}]%
\relax\mciteBstWouldAddEndPuncttrue
\mciteSetBstMidEndSepPunct{\mcitedefaultmidpunct}
{\mcitedefaultendpunct}{\mcitedefaultseppunct}\relax
\EndOfBibitem
\bibitem{H1:1996ovs}
S.~Aid et~al., The H1 collaboration, \emph{{Charged particle multiplicities in
  deep inelastic scattering at HERA}}, Z. Phys. C \textbf{72} (1996),
  \href{http://inspirehep.net/search?p=hep-ex/9608011}{573--592},
  [\href{http://arXiv.org/pdf/hep-ex/9608011}{{\texttt{hep-ex/9608011}}}]%
\relax\mciteBstWouldAddEndPuncttrue
\mciteSetBstMidEndSepPunct{\mcitedefaultmidpunct}
{\mcitedefaultendpunct}{\mcitedefaultseppunct}\relax
\EndOfBibitem
\bibitem{H1:1995cqf}
S.~Aid et~al., The H1 collaboration, \emph{{A Study of the fragmentation of
  quarks in $e^- p$ collisions at HERA}}, Nucl. Phys. B \textbf{445} (1995),
  \href{http://inspirehep.net/search?p=hep-ex/9505003}{3--21},
  [\href{http://arXiv.org/pdf/hep-ex/9505003}{{\texttt{hep-ex/9505003}}}]%
\relax\mciteBstWouldAddEndPuncttrue
\mciteSetBstMidEndSepPunct{\mcitedefaultmidpunct}
{\mcitedefaultendpunct}{\mcitedefaultseppunct}\relax
\EndOfBibitem
\bibitem{H1:1997mpq}
C.~Adloff et~al., The H1 collaboration, \emph{{Evolution of e p fragmentation
  and multiplicity distributions in the Breit frame}}, Nucl. Phys. B
  \textbf{504} (1997),
  \href{http://inspirehep.net/search?p=hep-ex/9707005}{3--23},
  [\href{http://arXiv.org/pdf/hep-ex/9707005}{{\texttt{hep-ex/9707005}}}]%
\relax\mciteBstWouldAddEndPuncttrue
\mciteSetBstMidEndSepPunct{\mcitedefaultmidpunct}
{\mcitedefaultendpunct}{\mcitedefaultseppunct}\relax
\EndOfBibitem
\bibitem{Dasgupta:2002dc}
M.~Dasgupta and G.~P. Salam, \emph{{Resummed event shape variables in DIS}},
  JHEP \textbf{08} (2002),
  \href{http://inspirehep.net/search?p=hep-ph/0208073}{032},
  [\href{http://arXiv.org/pdf/hep-ph/0208073}{{\texttt{hep-ph/0208073}}}]%
\relax\mciteBstWouldAddEndPuncttrue
\mciteSetBstMidEndSepPunct{\mcitedefaultmidpunct}
{\mcitedefaultendpunct}{\mcitedefaultseppunct}\relax
\EndOfBibitem
\bibitem{Stewart:2010tn}
I.~W. Stewart, F.~J. Tackmann and W.~J. Waalewijn, \emph{{N-Jettiness: An
  Inclusive Event Shape to Veto Jets}}, Phys. Rev. Lett. \textbf{105} (2010),
  \href{http://inspirehep.net/search?p=1004.2489}{092002},
  [\href{http://arXiv.org/pdf/1004.2489}{{\texttt{arXiv:1004.2489}}} [hep-ph]]%
\relax\mciteBstWouldAddEndPuncttrue
\mciteSetBstMidEndSepPunct{\mcitedefaultmidpunct}
{\mcitedefaultendpunct}{\mcitedefaultseppunct}\relax
\EndOfBibitem
\bibitem{Jouttenus:2011wh}
T.~T. Jouttenus, I.~W. Stewart, F.~J. Tackmann and W.~J. Waalewijn, \emph{{The
  Soft Function for Exclusive N-Jet Production at Hadron Colliders}}, Phys.
  Rev. D \textbf{83} (2011),
  \href{http://inspirehep.net/search?p=1102.4344}{114030},
  [\href{http://arXiv.org/pdf/1102.4344}{{\texttt{arXiv:1102.4344}}} [hep-ph]]%
\relax\mciteBstWouldAddEndPuncttrue
\mciteSetBstMidEndSepPunct{\mcitedefaultmidpunct}
{\mcitedefaultendpunct}{\mcitedefaultseppunct}\relax
\EndOfBibitem
\bibitem{Kang:2012zr}
Z.-B. Kang, S.~Mantry and J.-W. Qiu, \emph{{N-Jettiness as a Probe of Nuclear
  Dynamics}}, Phys. Rev. D \textbf{86} (2012),
  \href{http://inspirehep.net/search?p=1204.5469}{114011},
  [\href{http://arXiv.org/pdf/1204.5469}{{\texttt{arXiv:1204.5469}}} [hep-ph]]%
\relax\mciteBstWouldAddEndPuncttrue
\mciteSetBstMidEndSepPunct{\mcitedefaultmidpunct}
{\mcitedefaultendpunct}{\mcitedefaultseppunct}\relax
\EndOfBibitem
\bibitem{Kang:2013wca}
Z.-B. Kang, X.~Liu, S.~Mantry and J.-W. Qiu, \emph{{Probing nuclear dynamics in
  jet production with a global event shape}}, Phys. Rev. D \textbf{88} (2013),
  \href{http://inspirehep.net/search?p=1303.3063}{074020},
  [\href{http://arXiv.org/pdf/1303.3063}{{\texttt{arXiv:1303.3063}}} [hep-ph]]%
\relax\mciteBstWouldAddEndPuncttrue
\mciteSetBstMidEndSepPunct{\mcitedefaultmidpunct}
{\mcitedefaultendpunct}{\mcitedefaultseppunct}\relax
\EndOfBibitem
\bibitem{Kang:2014qba}
D.~Kang, C.~Lee and I.~W. Stewart, \emph{{Analytic calculation of 1-jettiness
  in DIS at $ \mathcal{O}\left({\alpha}_s\right) $}}, JHEP \textbf{11} (2014),
  \href{http://inspirehep.net/search?p=1407.6706}{132},
  [\href{http://arXiv.org/pdf/1407.6706}{{\texttt{arXiv:1407.6706}}} [hep-ph]]%
\relax\mciteBstWouldAddEndPuncttrue
\mciteSetBstMidEndSepPunct{\mcitedefaultmidpunct}
{\mcitedefaultendpunct}{\mcitedefaultseppunct}\relax
\EndOfBibitem
\bibitem{Kang:2013lga}
Z.-B. Kang, X.~Liu and S.~Mantry, \emph{{1-jettiness DIS event shape: NNLL+NLO
  results}}, Phys. Rev. D \textbf{90} (2014), no.~1,
  \href{http://inspirehep.net/search?p=1312.0301}{014041},
  [\href{http://arXiv.org/pdf/1312.0301}{{\texttt{arXiv:1312.0301}}} [hep-ph]]%
\relax\mciteBstWouldAddEndPuncttrue
\mciteSetBstMidEndSepPunct{\mcitedefaultmidpunct}
{\mcitedefaultendpunct}{\mcitedefaultseppunct}\relax
\EndOfBibitem
\bibitem{Kang:2015swk}
D.~Kang, C.~Lee and I.~W. Stewart, \emph{{DIS Event Shape at N3LL}}, PoS
  \textbf{DIS2015} (2015),
  \href{http://inspirehep.net/search?j=PoS,DIS2015,142}{142}%
\relax\mciteBstWouldAddEndPuncttrue
\mciteSetBstMidEndSepPunct{\mcitedefaultmidpunct}
{\mcitedefaultendpunct}{\mcitedefaultseppunct}\relax
\EndOfBibitem
\bibitem{Frye:2016aiz}
C.~Frye, A.~J. Larkoski, M.~D. Schwartz and K.~Yan, \emph{{Factorization for
  groomed jet substructure beyond the next-to-leading logarithm}}, JHEP
  \textbf{07} (2016), \href{http://inspirehep.net/search?p=1603.09338}{064},
  [\href{http://arXiv.org/pdf/1603.09338}{{\texttt{arXiv:1603.09338}}}
  [hep-ph]]%
\relax\mciteBstWouldAddEndPuncttrue
\mciteSetBstMidEndSepPunct{\mcitedefaultmidpunct}
{\mcitedefaultendpunct}{\mcitedefaultseppunct}\relax
\EndOfBibitem
\bibitem{Hoang:2019ceu}
A.~H. Hoang, S.~Mantry, A.~Pathak and I.~W. Stewart, \emph{{Nonperturbative
  Corrections to Soft Drop Jet Mass}}, JHEP \textbf{12} (2019),
  \href{http://inspirehep.net/search?p=1906.11843}{002},
  [\href{http://arXiv.org/pdf/1906.11843}{{\texttt{arXiv:1906.11843}}}
  [hep-ph]]%
\relax\mciteBstWouldAddEndPuncttrue
\mciteSetBstMidEndSepPunct{\mcitedefaultmidpunct}
{\mcitedefaultendpunct}{\mcitedefaultseppunct}\relax
\EndOfBibitem
\bibitem{Baberuxki:2019ifp}
N.~Baberuxki, C.~T. Preuss, D.~Reichelt and S.~Schumann, \emph{{Resummed
  predictions for jet-resolution scales in multijet production in
  e$^{+}$e$^{-}$ annihilation}}, JHEP \textbf{04} (2020),
  \href{http://inspirehep.net/search?p=1912.09396}{112},
  [\href{http://arXiv.org/pdf/1912.09396}{{\texttt{arXiv:1912.09396}}}
  [hep-ph]]%
\relax\mciteBstWouldAddEndPuncttrue
\mciteSetBstMidEndSepPunct{\mcitedefaultmidpunct}
{\mcitedefaultendpunct}{\mcitedefaultseppunct}\relax
\EndOfBibitem
\bibitem{Ritzmann:2014mka}
M.~Ritzmann and W.~J. Waalewijn, \emph{{Fragmentation in Jets at NNLO}}, Phys.
  Rev. D \textbf{90} (2014), no.~5,
  \href{http://inspirehep.net/search?p=1407.3272}{054029},
  [\href{http://arXiv.org/pdf/1407.3272}{{\texttt{arXiv:1407.3272}}} [hep-ph]]%
\relax\mciteBstWouldAddEndPuncttrue
\mciteSetBstMidEndSepPunct{\mcitedefaultmidpunct}
{\mcitedefaultendpunct}{\mcitedefaultseppunct}\relax
\EndOfBibitem
\bibitem{Caletti:2021oor}
S.~Caletti, O.~Fedkevych, S.~Marzani, D.~Reichelt, S.~Schumann, G.~Soyez and
  V.~Theeuwes, \emph{{Jet angularities in Z+jet production at the LHC}}, JHEP
  \textbf{07} (2021), \href{http://inspirehep.net/search?p=2104.06920}{076},
  [\href{http://arXiv.org/pdf/2104.06920}{{\texttt{arXiv:2104.06920}}}
  [hep-ph]]%
\relax\mciteBstWouldAddEndPuncttrue
\mciteSetBstMidEndSepPunct{\mcitedefaultmidpunct}
{\mcitedefaultendpunct}{\mcitedefaultseppunct}\relax
\EndOfBibitem
\bibitem{Caletti:2021ysv}
S.~Caletti, O.~Fedkevych, S.~Marzani and D.~Reichelt, \emph{{Tagging the
  initial-state gluon}}, Eur. Phys. J. C \textbf{81} (2021), no.~9,
  \href{http://inspirehep.net/search?p=2108.10024}{844},
  [\href{http://arXiv.org/pdf/2108.10024}{{\texttt{arXiv:2108.10024}}}
  [hep-ph]]%
\relax\mciteBstWouldAddEndPuncttrue
\mciteSetBstMidEndSepPunct{\mcitedefaultmidpunct}
{\mcitedefaultendpunct}{\mcitedefaultseppunct}\relax
\EndOfBibitem
\bibitem{Gleisberg:2007md}
T.~Gleisberg and F.~Krauss, \emph{{Automating dipole subtraction for QCD NLO
  calculations}}, Eur. Phys. J. C \textbf{53} (2008),
  \href{http://inspirehep.net/search?p=0709.2881}{501--523},
  [\href{http://arXiv.org/pdf/0709.2881}{{\texttt{arXiv:0709.2881}}} [hep-ph]]%
\relax\mciteBstWouldAddEndPuncttrue
\mciteSetBstMidEndSepPunct{\mcitedefaultmidpunct}
{\mcitedefaultendpunct}{\mcitedefaultseppunct}\relax
\EndOfBibitem
\bibitem{Actis:2016mpe}
S.~Actis, A.~Denner, L.~Hofer, J.-N. Lang, A.~Scharf and S.~Uccirati,
  \emph{{RECOLA: REcursive Computation of One-Loop Amplitudes}}, Comput. Phys.
  Commun. \textbf{214} (2017),
  \href{http://inspirehep.net/search?p=1605.01090}{140--173},
  [\href{http://arXiv.org/pdf/1605.01090}{{\texttt{arXiv:1605.01090}}}
  [hep-ph]]%
\relax\mciteBstWouldAddEndPuncttrue
\mciteSetBstMidEndSepPunct{\mcitedefaultmidpunct}
{\mcitedefaultendpunct}{\mcitedefaultseppunct}\relax
\EndOfBibitem
\bibitem{Biedermann:2017yoi}
B.~Biedermann, S.~Br\"auer, A.~Denner, M.~Pellen, S.~Schumann and J.~M.
  Thompson, \emph{{Automation of NLO QCD and EW corrections with Sherpa and
  Recola}}, Eur. Phys. J. C \textbf{77} (2017),
  \href{http://inspirehep.net/search?p=1704.05783}{492},
  [\href{http://arXiv.org/pdf/1704.05783}{{\texttt{arXiv:1704.05783}}}
  [hep-ph]]%
\relax\mciteBstWouldAddEndPuncttrue
\mciteSetBstMidEndSepPunct{\mcitedefaultmidpunct}
{\mcitedefaultendpunct}{\mcitedefaultseppunct}\relax
\EndOfBibitem
\bibitem{Cascioli:2011va}
F.~Cascioli, P.~Maierhofer and S.~Pozzorini, \emph{{Scattering Amplitudes with
  Open Loops}}, Phys. Rev. Lett. \textbf{108} (2012),
  \href{http://inspirehep.net/search?p=1111.5206}{111601},
  [\href{http://arXiv.org/pdf/1111.5206}{{\texttt{arXiv:1111.5206}}} [hep-ph]]%
\relax\mciteBstWouldAddEndPuncttrue
\mciteSetBstMidEndSepPunct{\mcitedefaultmidpunct}
{\mcitedefaultendpunct}{\mcitedefaultseppunct}\relax
\EndOfBibitem
\bibitem{Dobbs:2001ck}
M.~Dobbs and J.~B. Hansen, \emph{{The HepMC C++ Monte Carlo event record for
  High Energy Physics}}, Comput. Phys. Commun. \textbf{134} (2001),
  \href{http://inspirehep.net/search?j=Comput%20Phys%20Commun,134,41}{41--46}%
\relax\mciteBstWouldAddEndPuncttrue
\mciteSetBstMidEndSepPunct{\mcitedefaultmidpunct}
{\mcitedefaultendpunct}{\mcitedefaultseppunct}\relax
\EndOfBibitem
\bibitem{Dasgupta:2020fwr}
M.~Dasgupta, F.~A. Dreyer, K.~Hamilton, P.~F. Monni, G.~P. Salam and G.~Soyez,
  \emph{{Parton showers beyond leading logarithmic accuracy}}, Phys. Rev. Lett.
  \textbf{125} (2020), no.~5,
  \href{http://inspirehep.net/search?p=2002.11114}{052002},
  [\href{http://arXiv.org/pdf/2002.11114}{{\texttt{arXiv:2002.11114}}}
  [hep-ph]]%
\relax\mciteBstWouldAddEndPuncttrue
\mciteSetBstMidEndSepPunct{\mcitedefaultmidpunct}
{\mcitedefaultendpunct}{\mcitedefaultseppunct}\relax
\EndOfBibitem
\bibitem{Forshaw:2020wrq}
J.~R. Forshaw, J.~Holguin and S.~Pl\"atzer, \emph{{Building a consistent parton
  shower}}, JHEP \textbf{09} (2020),
  \href{http://inspirehep.net/search?p=2003.06400}{014},
  [\href{http://arXiv.org/pdf/2003.06400}{{\texttt{arXiv:2003.06400}}}
  [hep-ph]]%
\relax\mciteBstWouldAddEndPuncttrue
\mciteSetBstMidEndSepPunct{\mcitedefaultmidpunct}
{\mcitedefaultendpunct}{\mcitedefaultseppunct}\relax
\EndOfBibitem
\bibitem{Herren:2022jej}
\href{http://inspirehep.net/search?p=2208.06057}{F.~Herren, S.~H\"oche,
  F.~Krauss, D.~Reichelt and M.~Schoenherr}, \emph{{A new approach to
  color-coherent parton evolution}},
  \href{http://arXiv.org/pdf/2208.06057}{{\texttt{arXiv:2208.06057}}} [hep-ph]%
\relax\mciteBstWouldAddEndPuncttrue
\mciteSetBstMidEndSepPunct{\mcitedefaultmidpunct}
{\mcitedefaultendpunct}{\mcitedefaultseppunct}\relax
\EndOfBibitem
\bibitem{vanBeekveld:2023lfu}
\href{http://inspirehep.net/search?p=2305.08645}{M.~van Beekveld and
  S.~Ferrario~Ravasio}, \emph{{Next-to-leading-logarithmic PanScales showers
  for Deep Inelastic Scattering and Vector Boson Fusion}},
  \href{http://arXiv.org/pdf/2305.08645}{{\texttt{arXiv:2305.08645}}} [hep-ph]%
\relax\mciteBstWouldAddEndPuncttrue
\mciteSetBstMidEndSepPunct{\mcitedefaultmidpunct}
{\mcitedefaultendpunct}{\mcitedefaultseppunct}\relax
\EndOfBibitem
\bibitem{Hoche:2017kst}
S.~H\"oche, D.~Reichelt and F.~Siegert, \emph{{Momentum conservation and
  unitarity in parton showers and NLL resummation}}, JHEP \textbf{01} (2018),
  \href{http://inspirehep.net/search?p=1711.03497}{118},
  [\href{http://arXiv.org/pdf/1711.03497}{{\texttt{arXiv:1711.03497}}}
  [hep-ph]]%
\relax\mciteBstWouldAddEndPuncttrue
\mciteSetBstMidEndSepPunct{\mcitedefaultmidpunct}
{\mcitedefaultendpunct}{\mcitedefaultseppunct}\relax
\EndOfBibitem
\bibitem{H1:1996muf}
C.~Adloff et~al., The H1 collaboration, \emph{{Measurement of charged particle
  transverse momentum spectra in deep inelastic scattering}}, Nucl. Phys. B
  \textbf{485} (1997),
  \href{http://inspirehep.net/search?p=hep-ex/9610006}{3--24},
  [\href{http://arXiv.org/pdf/hep-ex/9610006}{{\texttt{hep-ex/9610006}}}]%
\relax\mciteBstWouldAddEndPuncttrue
\mciteSetBstMidEndSepPunct{\mcitedefaultmidpunct}
{\mcitedefaultendpunct}{\mcitedefaultseppunct}\relax
\EndOfBibitem
\end{mcitethebibliography}

\end{document}